# VIBES-A Two-Stage Scalable Bayesian Uncertainty Quantification Framework: Application to a Biomass Valorization Process

Poulomi Das[1], Angan Mukherjee[2], and Debangsu Bhattacharyya[1]*

[1]Department of Chemical and Biomedical Engineering, West Virginia University, Morgantown, WV 26506, USA
[2]Department of Chemical and Biological Engineering, University of Wisconsin-Madison, Madison, WI 53706, USA

## Abstract

This paper proposes Variational Inference-based Bayesian Estimation with Sobol screening (VIBES) – a two-stage scalable framework for Bayesian uncertainty quantification (UQ). The proposed approach combines Sobol global sensitivity analysis (GSA) for screening and dimensionality reduction, followed by variational inference (VI) for UQ of kinetic, design/operational, and economic parameters. In the first stage, Sobol GSA is performed to identify dominant variables and parameters governing uncertainty in process outputs. In the second stage, Bayesian inference is performed only on the reduced dimensional space using VI, thus reducing computational burden and enhancing scalability. The framework is demonstrated on a process for bioadhesive production through base-catalyzed depolymerization of kraft lignin and subsequent crosslinking with isolated soy protein. A Python-Aspen interface is developed for automated simulation and parameter estimation, enabling Bayesian calibration through stochastic gradient-based optimization and automatic-differentiation. The methodology is generic and readily generalizable to other biomass conversion pathways. The results show that application of VIBES consistently reduces predictive uncertainty bounds across all model outputs by more than 80%, even when only the reduced-space input variables and parameters are optimized during Bayesian estimation. The framework can be potentially applied for scalable, uncertainty-aware decision-making in high-dimensional, complex chemical process systems.

---

*Corresponding author. Tel: +1-3042939335, Fax: +1-3042934139

E-mail address: Debangsu.Bhattacharyya@mail.wvu.edu

## 1. Introduction

High-dimensional process models of chemical process systems can be used for process design, techno-economic analysis, sustainability analysis, and many other objectives. Such models are developed using mass, energy, and momentum conservation equations that might require sub-models for reaction kinetics, thermodynamics, physical property, heat and mass transport, capital and operating cost estimates, etc. However, there are often uncertainties in the parameters and form of these sub-models that can lead to uncertainties in outputs from these models. Key outputs of interest often include both performance measures such as conversion, product yield, selectivity, utility consumption, emissions, etc., and economic metrics like net present value, minimum selling price, levelized cost of production, etc. (Monteiro and Bhattacharyya, 2026; Morgan et al., 2015). Therefore, uncertainty quantification (UQ) is desired beyond nominal deterministic simulations (Kennedy and O'Hagan, 2001; Morgan et al., 2015), especially when process models are used for design, analysis, and decision-making.

UQ is particularly challenging when multiple sources of uncertainty interact across scales and process subsystems. For example, parameter uncertainty can arise from limited kinetic, thermodynamic, and mass/heat transfer data availability, whereas model-form uncertainty can result from simplified mechanisms, empirical correlations, incomplete physical descriptions, or assumptions introduced during model development (Mukherjee et al., 2025b; Ostace et al., 2020). In addition, uncertainty in operating conditions, feed composition, experimental measurements, and boundary conditions can affect both model calibration and process-level predictions (Morgan et al., 2015; Ostace et al., 2020). It has been observed that uncertainty in property models, mass transfer correlations, reaction kinetics, and other constitutive sub-models can lead to appreciable uncertainty in process performance, design requirements, and scale-up decisions (Hughes et al., 2022; Morgan et al., 2018; Soares Chinen et al., 2018). UQ has been undertaken for various chemical engineering applications, including $CO_2$ adsorption process models (Kalyanaraman et al., 2016), hydraulic and mass transfer sub-models (Soares Chinen et al., 2018) and physical properties and thermodynamic models (Morgan et al., 2015; Ostace et al., 2020) for solvent-based carbon capture processes, crude distillation unit (Minh et al., 2018), and fluidized-based gasification systems (Pan and Pandey, 2016).

Despite these advances, the computational scalability of UQ remains a major challenge for nonlinear and high-dimensional process models (Ostace et al., 2020). Most UQ analyses involve uncertainty propagation through process models, life-cycle models, or supply-chain models with sampling approaches such as several variants of Monte Carlo type methods, Latin hypercube sampling, and other approaches (Ou et al., 2018; Shi and Guest, 2020; Vicari et al., 2012). Computational expense for these approaches increases rapidly when the uncertain input space includes a large number of kinetic, design/operating, and economic variables and parameters, particularly when propagation of each sample requires the solution of a rigorous process model. This challenge is further exacerbated when uncertainty needs to be propagated through complex, highly nonlinear, and large process models with comprehensive mass/heat integration, and economic calculations. Consequently, state-of-the-art UQ studies are often restricted to a lower dimensional input space often down-selected using local sensitivity measures, or use simplified surrogate representations, potentially overlooking nonlinear effects and interactions among uncertain quantities (Hughes et al., 2022; Kirchner et al., 2021; Strunge et al., 2023; Zhang, 2021).

Bayesian approaches to UQ are being increasingly investigated for a variety of applications. Bayesian approaches combine prior knowledge with available observations (measurements) to estimate posterior distributions of uncertain variables and/or parameters (Kennedy and O'Hagan, 2001). Additionally, these approaches can quantify the relevant ranges of model input variables and parameters that remain consistent with available process, experimental, or market-derived information. However, sampling-based Bayesian approaches, such as Markov chain Monte Carlo (MCMC), can become computationally expensive for high-dimensional nonlinear process models that require repeated first-principles model evaluations (Bardsley and Hansen, 2020; Zhang, 2021). For keeping computational expenses tractable, down-selection of the input space is often undertaken through local sensitivity analysis mostly around nominal operating conditions, but it may not adequately represent nonlinearities, higher-order interactions, or the global uncertainty structure spanning the combined input-output space (Strunge et al., 2023; Saltelli et al., 2010, 2019). These limitations motivate the development of scalable computational approaches for the unified and efficient end-to-end integration of global sensitivity analysis, reduced-dimensional Bayesian estimation, and posterior predictive UQ.

The need for such approaches is especially pronounced when UQ is considered from a techno-economic analysis (TEA) perspective. Specifically, the simultaneous dependence of TEA outputs on uncertain kinetic, design/operating, and economic variables and parameters (Das and Bhattacharyya, 2026a) leads to a high-dimensional and highly nonlinear inference problem. Uncertainty analysis of economic performance measures has also been reported. For example, propagation of uncertainty in feedstock properties, yields, capital costs, and market variables through Aspen-based TEA models (Vicari et al., 2012) or biorefinery simulation platforms such as BioSTEAM (Cortes-Peña et al., 2020; Shi and Guest, 2020) has been undertaken to generate distributions of product cost, minimum selling price, etc. While such approaches are useful for quantifying how assumed input uncertainty affects model predictions, they do not infer posterior distributions of uncertain variables and parameters from available experimental or market data (Bardsley and Hansen, 2020; Salam et al., 2024). In techno-economic studies, uncertainties in feedstock and product prices, utility tariffs, capital-cost factors, and policy variables are commonly examined through one-at-a-time sensitivity or scenario analysis, rather than through data-conditioned posterior estimation (Strunge et al., 2023; Van Der Spek et al., 2020). Even recent Bayesian optimization-based process economic studies use probabilistic surrogate models to identify favorable operating or design conditions and subsequently assess the effects of economic variables, but they do not seek joint posterior distributions of economic and process variables and parameters from process and market target data (Park et al., 2026; Urm et al., 2023). Consequently, there remains a need for developing an integrated framework that can sequentially identify dominant sources of uncertainty in process modeling and TEA workflows, followed by posterior estimation and Bayesian UQ in a computationally tractable reduced space.

To address these challenges, this paper proposes Variational Inference-based Bayesian Estimation with Sobol screening (VIBES), a two-stage scalable Bayesian framework for UQ. In the first stage, Sobol global sensitivity analysis (GSA) is undertaken to screen the full uncertain input space and identify dominant kinetic, design/operating, and economic variables as well as parameters that govern variability in process and economic outputs (Saltelli et al., 2010). This variance-based screening provides a rigorous dimensionality-reduction strategy that can still capture dominant nonlinear effects and parameter/variable interactions, while improving identifiability and computational tractability for the subsequent inference step. In the second stage, variational inference (VI) is performed over the reduced Sobol-selected input space to estimate posterior distributions of the influential uncertain variables and parameters (Mukherjee and Zavala, 2026; Wang et al., 2024). By reformulating posterior inference as an optimization problem through approximation of the true posterior distribution (Mukherjee and Zavala, 2026; Thompson et al., 2026; Wang et al., 2023), VI provides a scalable alternative to conventional Bayesian sampling methods, especially for computationally expensive models where repeated model evaluations are expensive and rapid uncertainty updates are essential for design-space exploration. Estimated posterior distributions are then propagated through the process model to quantify posterior predictive uncertainty in both process and economic outputs.

The proposed framework is generic and can be extended to high-dimensional chemical process systems, in general. In this work, VIBES is applied to a biomass valorization process. In recent years, lignocellulosic biomass valorization has emerged as an important route for the production of fuels, chemicals, and value-added materials with improved sustainability by reducing dependence on fossil-derived resources (Das and Bhattacharyya, 2026a;(Das and Bhattacharyya, 2026a; Kirchner et al., 2021). However, the commercial deployment of novel biomass conversion technologies remains significantly challenging due to several interacting sources of uncertainty that can have strong impact on process performance, product quality, production rate, and economic viability. These uncertainties may arise from feedstock composition, limited experimental data, uncertain reaction pathways and kinetic parameters, scale-up assumptions, and market-dependent techno-economic factors such as raw material prices, utility costs, and equipment cost correlations, to name a few (Das and Bhattacharyya, 2026b; Ou et al., 2018; Strunge et al., 2023). As a result, UQ of novel biomass conversion technologies is highly desired to support uncertainty-aware process design and decision-making for different biomass valorization pathways.

In this work, the VIBES framework is demonstrated using a bioadhesive production process involving base-catalyzed depolymerization (BCD) of kraft lignin followed by crosslinking with soy protein isolate (SPI) (Das and Bhattacharyya, 2026b; Jiang et al., 2023). The kraft lignin-soy protein isolate (KL-SPI) bioadhesive production technology is very novel and suffers from high uncertainties. In this work, available literature and market information are used to define physically meaningful ranges for selected kinetic, design/operating, and economic variables and parameters. These ranges provide representative information to generate synthetic data in the absence of sufficient experimental, pilot-plant, and/or market-derived observations for this particular or similar bio-adhesive manufacturing technologies. Obviously, any synthetic data that is used in this work can readily be replaced by measured or experimental data when such information becomes available. It is desired to investigate if the variability observed in target outputs, such as D-lignin (i.e., partially degraded lignin/oligomers (Das and Bhattacharyya, 2026b)) flow rate as a process performance measure and levelized cost of adhesive (LCOA) as an economic performance indicator, can be explained and quantified in terms of kinetic, design/operating, and economic uncertainties in the underlying input space. VIBES uses available process and economic target information to infer posterior distributions of influential variables and parameters. The resulting posterior distributions are then propagated through the process model to enable uncertainty-aware design-space exploration, recommendation of future experimental needs, and more reliable evaluation of KL-SPI bioadhesive manufacturing process.

Selecting appropriate software platform(s) is often an important decision for Bayesian UQ. While process models are often developed in a software that readily offers physical properties models and parameters and comprehensive library of process models of desired level of rigor for typical unit operations involved in chemical processes, tools/capabilities for UQ (Thompson et al., 2026), optimization (Mukherjee et al., 2025a; Tian et al., 2023), machine learning (ML) (Lin, 2024), and data analytics (Oladokun et al., 2022) may be more readily available or more efficiently implemented in another computational ecosystem (Nikkhah et al., 2024). When multiple software platforms are involved, it is important to consider compatibilities of these software, what information is needed vs what information is available to be communicated from one software to the other, and communication overhead. In this work, the process model for the bioadhesive production process is developed in Aspen Plus while other capabilities/tools for Bayesian UQ are developed in Python. Therefore, a multi-software integration approach is developed that enables automated input sampling, process simulation, output extraction, GSA, Bayesian inference, model evaluation, and posterior uncertainty propagation. This multi-software interface within VIBES further enables

Bayesian calibration through stochastic gradient-based optimization (Tian et al., 2023) and automatic differentiation using JAX (Lin, 2024) during iterative inference, while the high-fidelity process model is used to evaluate posterior predictive uncertainty. In addition, this work systematically compares and contrasts different dimensionalities of the Sobol-selected input space that achieves significant reduction in predictive uncertainty, rather than selecting a fixed reduced dimension *a priori*, thereby enhancing computational scalability during posterior estimation.

In summary, the main contributions of this paper are as follows:

- Development of VIBES, a two-stage scalable Bayesian framework that integrates Sobol GSA with VI for simultaneous estimation and UQ of dominant kinetic, design/operating, and economic variables and parameters in biomass valorization systems.
- Implementation of a sensitivity-driven dimensionality-reduction strategy that identifies the most influential variables and parameters governing uncertainty in process and economic outputs, thus improving inferential tractability, reducing over-parameterization, and enhancing scalability during posterior estimation.
- Development of a scalable posterior estimation and predictive UQ workflow, where VI is used to estimate posterior distributions of the Sobol-selected variables and parameters, followed by propagation of these distributions through the process model to quantify posterior predictive uncertainty in process performance and economic indicators.
- Development of a multi-software computational platform for automated simulation, parameter estimation, model evaluation, and uncertainty propagation, enabling Bayesian calibration through stochastic gradient-based optimization and automatic differentiation (AD) during iterative inference.
- Demonstration of the VIBES framework on a KL-SPI bioadhesive production process, where uncertainties in D-lignin mass flow rate and LCOA are quantified due to uncertainties in interacting kinetic, design/operating, and economic variables and parameters, thus resulting in uncertainty-aware design-space exploration and reliable decision-making for biomass valorization technologies.

The rest of the paper is organized as follows. Section 2 presents the detailed computational workflow of VIBES, including the formulation of the reduced dimensional input space with Sobol-screened variables and parameters, evaluation of posterior and predictive uncertainty using VI for Bayesian inference, followed by the development of a multi-software integration framework for automated simulation and analysis. Section 3 describes the KL-SPI bioadhesive production process, along with the relevant implementation details of the multi-software interface and the list of uncertain input variables and parameters considered in this work. Section 4 presents and discusses the results obtained from the implementation of VIBES for simultaneous parameter estimation and UQ in the selected case study. Finally, Section 5 summarizes the major findings in the concluding remarks and presents potential directions for future work.

## 2. VIBES: Computational Workflow

This section presents the overall computational framework of VIBES. The proposed workflow leads to the formulation of a two-stage Bayesian approach that integrates Sobol GSA with VI for scalable parameter estimation and UQ. The separation of sensitivity-driven dimensionality reduction and Bayesian estimation enhances the scalability of VIBES. The framework can be implemented entirely within a single computational ecosystem or through a multi-software integration platform, as discussed in Section 2.3.

### 2.1. Sobol Global Sensitivity Analysis

The first stage of VIBES employs Sobol GSA to identify the subset of dominating uncertain variables and parameters for the process and economic outputs. The objective of this screening step is to reduce the dimensionality of the Bayesian estimation (optimization) problem before applying VI. It is worth mentioning here that VIBES solves an inverse problem for estimating uncertainty in the uncertain variables and parameters given available experimental or market-derived data. Simultaneously estimating all uncertain input variables and parameters would not only lead to high computational expense, but also can lead to poor estimability especially when availability of experimental or market-derived data is scarce as would be expected for novel biomass valorization technologies. This further motivates the sensitivity-guided dimensionality reduction before Bayesian inference.

Let the process model be represented as $\boldsymbol{y} = f(\boldsymbol{\theta})$, where $\boldsymbol{\theta} = [\theta_1, \theta_2, \dots, \theta_D]^T \in \mathbb{R}^D$ denotes the vector of $D$ uncertain input variables and parameters. The corresponding model output vector can be defined as

$$\boldsymbol{y} = [y_1, y_2, \dots, y_M] \in \mathbb{R}^M \tag{1}$$

where each output $y_m$ for $m = 1,2, \dots, M$ may represent a process performance and/or economic indicator of interest. For each uncertain variable or parameter $\theta_j$, the respective lower and upper bounds are specified as

$$\theta_j = [\theta_j^L, \ \ \theta_j^U], \quad j = 1, \dots, D \tag{2}$$

where $\theta_j^L$ and $\theta_j^U$ refer to the lower and upper bounds, respectively, of the $j^{th}$ uncertain variable / parameter. These bounds define the joint uncertainty domain over which the Sobol screening is performed.

For each model output $y_m$, the total-order Sobol index of the $j^{th}$ uncertain variable or parameter is defined as

$$S_{T_j}(y_m) = 1 - \frac{Var\left(\mathbb{E}\left[y_m \mid \boldsymbol{\theta}_{\sim j}\right]\right)}{Var\,(y_m)} \tag{3}$$

where $\boldsymbol{\theta}_{\sim j}$ represents the vector of all uncertain variables and parameters except $\theta_j$ for $Var(y_m) > 0$. The total-order index $S_{T_j}$ measures the total contribution of $\theta_j$ to the variance of the output $y_m$, including both its direct effect and all higher-order interaction effects with other variables and parameters (Puy et al., 2022; Tosin et al., 2020; Zhang et al., 2015). The total-order Sobol index accounts for nonlinear interactions among various model inputs.

In practice, the Sobol indices are estimated using a Sobol quasi-Monte Carlo sampling strategy (Saltelli et al., 2010). First, $2N$ samples are generated in the normalized unit hypercube $[0, 1]^D$ and split into two base matrices defined as

$$U_A \in [0, 1]^{N \times D}, \quad U_B \in [0, 1]^{N \times D} \tag{4}$$

where $N$ denotes the number of samples in each base matrix. These normalized samples are then mapped to the physical input bounds as

$$A_{n,j} = \theta_j^L + U_{A_{n,j}} \left(\theta_j^U - \theta_j^L\right) \tag{5a}$$

$$B_{n,j} = \theta_j^L + U_{B_{n,j}} \left(\theta_j^U - \theta_j^L\right) \tag{5b}$$

for sample index $n = 1,2, \dots, N$ and input index $j = 1,2, \dots, D$. Here, $A$ and $B$ are the two physical sample matrices corresponding to $U_A$ and $U_B$. To estimate the contribution of each uncertain variable or parameter, hybrid matrices are constructed by replacing one column of $A$ at a time with the corresponding column of $B$. For the $j^{th}$ uncertain variable or parameter, the hybrid matrix is defined as

$$AB^{(j)} = A \text{ with column } j \text{ replaced by the } j^{th} \text{ column of } B \tag{6}$$

Thus, $AB^{(j)}$ differs from $A$ only in the $j^{th}$ input variable or parameter. The process model is then evaluated for the rows of $A, B,$ and each hybrid matrix $AB^{(j)}$. Therefore, for a given output $y_m$, the corresponding model evaluations can be expressed as

$$Y_{A,m}^{(n)} = f_m\left(A_{n,:}\right), \quad n = 1, 2, \dots, N \tag{7a}$$

$$Y_{B,m}^{(n)} = f_m\left(B_{n,:}\right), \quad n = 1, 2, \dots, N \tag{7b}$$

$$Y_{AB^{(j)},m}^{(n)} = f_m\left(AB_{n,:}^{(j)}\right), \quad n = 1, 2, \dots, N \tag{7c}$$

where $f_m(\cdot)$ refers to the $m^{th}$ scalar model output, and $A_{n,:}, B_{n,:}$, and $AB_{n,:}^{(j)}$ denote the $n^{th}$ rows of the corresponding sample matrices. Here, $Y_{A,m}^{(n)}, Y_{B,m}^{(n)}$, and $Y_{AB^{(j)},m}^{(n)}$ are the scalar model evaluations for the $n^{th}$ sample, while the corresponding output vectors are defined as

$$\mathbf{Y}_{A,m} = \left[Y_{A,m}^{(1)}, \;\; Y_{A,m}^{(2)}, \dots, \;\; Y_{A,m}^{(N)}\right]^T \tag{8a}$$

$$\mathbf{Y}_{B,m} = \left[Y_{B,m}^{(1)}, \;\; Y_{B,m}^{(2)}, \dots, \;\; Y_{B,m}^{(N)}\right]^T \tag{8b}$$

$$\mathbf{Y}_{AB^{(j)},m} = \left[Y_{AB^{(j)},m}^{(1)}, \;\; Y_{AB^{(j)},m}^{(2)}, \dots, \;\; Y_{AB^{(j)},m}^{(N)}\right]^T \tag{8c}$$

The output variance for $y_m$ is then estimated using the combined model evaluations from the two base matrices, given by

$$V_{y_m} = Var\left(\left[\mathbf{Y}_{A,m};\ \mathbf{Y}_{B,m}\right]\right) \tag{9}$$

Finally, using the Saltelli estimator (Saltelli et al., 2019, 2010), the total-order Sobol index for the $j^{th}$ uncertain variable or parameter with respect to output $y_m$ is estimated as

$$\hat{S}_{T_j}(y_m) = \frac{1}{2N} * \frac{1}{V_{y_m}} * \sum_{n=1}^{N} \left(Y_{A,m}^{(n)} - Y_{AB^{(j)},m}^{(n)}\right)^2 \tag{10}$$

This estimator, as defined in Eq. (10), quantifies how strongly the output variance changes when the $j^{th}$ uncertain input is perturbed within its admissible range while accounting for the interaction effects with the remaining variables and parameters.

In this work, since the process model produces multiple outputs of interest, VIBES ranks the uncertain variables and parameters based on their average total-order Sobol contribution across all selected process and/or economic outputs. Specifically, the average total-order Sobol index for the $j^{th}$ uncertain variable or parameter can be computed as

$$\bar{S}_{T_j} = \frac{1}{M} \sum_{m=1}^{M} \hat{S}_{T_j}(y_m) \tag{11}$$

where $M$ is the number of outputs considered in the screening step. The dominant variables and parameters are then selected according to the largest values of $\bar{S}_{T_j}$. For a chosen reduced-space dimension $N_{top}$, the selected index set is defined as

$$\boldsymbol{\mathcal{I}}_{N_{top}} = \underset{N_{top}}{\operatorname{argtop}} \left\{\bar{S}_{T_j}\right\}_{j=1}^{D} \tag{12}$$

where $\mathcal{I}_{N_{top}}$ denotes the indices corresponding to the $N_{top}$ largest average total-order Sobol indices. The resulting reduced input vector is then expressed as

$$\boldsymbol{\theta}_{N_{top}} = \left\{\theta_j : j \in \boldsymbol{\mathcal{I}}_{N_{top}}\right\} \tag{13}$$

It should be noted that VI is undertaken only for the selected subset $\boldsymbol{\theta}_{N_{top}}$. In this work, different values of $N_{top}$ are also systematically evaluated to determine how the dimension of the Sobol-selected input space affects posterior contraction and subsequent reduction in predictive uncertainty.

### 2.2. Variational Inference

The second stage of VIBES performs Bayesian estimation and posterior UQ over the reduced input space obtained from the Sobol GSA screening step. As defined in Eq. (13), the Sobol-selected reduced input vector is represented as $\boldsymbol{\theta}_{N_{top}}$ that corresponds to $N_{top}$ variables and parameters with the largest average total-order Sobol indices $\left(\bar{S}_{T_j}\right)$. The objective of VI is to approximate the posterior distribution $p\left(\boldsymbol{\theta}_{N_{top}} \mid \boldsymbol{\mathcal{D}}\right)$, where $\boldsymbol{\mathcal{D}}$ denotes the available measurement or target data used for Bayesian estimation. Thus, VI estimates the posterior distributions of the selected variables and parameters that define regions of the design space capable of predicting outputs consistent with available experimental or market-derived data. Furthermore, unlike conventional sampling-based Bayesian UQ approaches, such as MCMC, that can be computationally intractable for nonlinear process models requiring repeated model evaluations, VI provides a scalable alternative by reformulating posterior estimation as an optimization problem.

Specifically, the posterior distribution is approximated using a parameterized variational distribution $q\left(\boldsymbol{\theta}_{N_{top}} \mid \boldsymbol{\phi}\right)$, where $\boldsymbol{\phi}$ denotes the variational parameters. The optimal variational parameters $(\boldsymbol{\phi}^*)$ are obtained by minimizing the Kullback-Leibler (KL) divergence (Blei et al., 2017; Thompson et al., 2026) between the variational approximation and the true posterior, given by

$$\boldsymbol{\phi}^* = \underset{\boldsymbol{\phi}}{\operatorname{argmin}} \ KL\left[q\left(\boldsymbol{\theta}_{N_{top}} \mid \boldsymbol{\phi}\right) || \, p\left(\boldsymbol{\theta}_{N_{top}} \mid \boldsymbol{\mathcal{D}}\right)\right] \tag{14}$$

where

$$KL\left[q\left(\boldsymbol{\theta}_{N_{top}} \mid \boldsymbol{\phi}\right) || \, p\left(\boldsymbol{\theta}_{N_{top}} \mid \boldsymbol{\mathcal{D}}\right)\right] = \int q\left(\boldsymbol{\theta}_{N_{top}} \mid \boldsymbol{\phi}\right) \log \left[\frac{q\left(\boldsymbol{\theta}_{N_{top}} \mid \boldsymbol{\phi}\right)}{p\left(\boldsymbol{\theta}_{N_{top}} \mid \boldsymbol{\mathcal{D}}\right)}\right] d\boldsymbol{\theta}_{N_{top}} \tag{15}$$

The KL-divergence minimization formulation as described in Eq. (14) can be equivalently expressed as negative evidence lower bound (ELBO) objective comprising three competing contributions. Specifically, the VI objective function can be defined as

$$J(\boldsymbol{\phi}) = \mathbb{E}_q\left[-\log\, p\left(\boldsymbol{\mathcal{D}} \mid \boldsymbol{\theta}_{N_{top}}\right)\right] - \mathbb{E}_q\left[\log\, p\left(\boldsymbol{\theta}_{N_{top}}\right)\right] - H\left[q\left(\boldsymbol{\theta}_{N_{top}} \mid \boldsymbol{\phi}\right)\right] \tag{16}$$

where the entropy of the variational distribution is represented as

$$H\left[q\left(\boldsymbol{\theta}_{N_{top}} \mid \boldsymbol{\phi}\right)\right] = -\mathbb{E}_q\left[\log\, q\left(\boldsymbol{\theta}_{N_{top}} \mid \boldsymbol{\phi}\right)\right] \tag{17}$$

The entropy, as defined in Eq. (17), preserves uncertainty in regions of the input space where limited or no experimental / market data are available, thus allowing broader predictive spreads when the data do not strongly constrain the posterior (Blei et al., 2017; Wang et al., 2023).

Therefore, the VI objective function as defined in Eq. (16), can be updated as

$$J(\boldsymbol{\phi}) = \mathbb{E}_q\left[-\log\, p\left(\boldsymbol{\mathcal{D}} \mid \boldsymbol{\theta}_{N_{top}}\right)\right] + \mathbb{E}_q\left[-\log\, p\left(\boldsymbol{\theta}_{N_{top}}\right)\right] + \mathbb{E}_q\left[\log\, q\left(\boldsymbol{\theta}_{N_{top}} \mid \boldsymbol{\phi}\right)\right] \tag{18}$$

The first term in Eq. (18) minimizes the error between model predictions and available process/economic target data, the second term penalizes deviation from the prior distribution, and the third term maximizes the entropy of the variational distribution while promoting a non-degenerate posterior approximation. Thus, VI enables simultaneous estimation and UQ of the Sobol-selected parameters and variables.

In this work, a diagonal Gaussian variational distribution is considered for the Sobol-selected variables and parameters. For each selected variable or parameter $\theta_j$, the reparameterized representation is defined as

$$\theta_j = \mu_j + \sigma_j\, z_j, \qquad j = 1, 2, \ldots, N_{top} \tag{19a}$$

$$z_j \sim \mathcal{N}(0,1) \tag{19b}$$

where $\mu_j$ and $\sigma_j$ are the variational mean and standard deviation associated with the $j^{th}$ selected variable or parameter, and $z_j$ is a standard normal (Gaussian) random variable. Therefore, the variational parameters are given by $\boldsymbol{\phi} = [\boldsymbol{\mu}, \boldsymbol{\sigma}]$. To improve the numerical stability of VI, a *warm-start initialization* strategy is implemented. Let the bank of successful model evaluations generated during the Sobol GSA screening stage be denoted as

$$\boldsymbol{\mathcal{D}}^{*} = \left\{ \left( \boldsymbol{\theta}_{N_{top}}^{(r)}, \boldsymbol{y}^{(r)} \right) \right\}_{r=1}^{N_{bank}} \tag{20}$$

where $N_{bank}$ represents the number of successful model simulations available from the Sobol sampling stage, $\boldsymbol{\theta}_{N_{top}}^{(r)}$ contains the Sobol-selected variables and parameters for the $r^{th}$ successful model simulation, and $\boldsymbol{y}^{(r)}$ denotes the corresponding simulated process and economic output vectors. Subsequently, the output misfit for each stored sample can be computed as

$$\mathcal{E}^{(r)} = \sum_{m=1}^{M} \left\| \boldsymbol{y}_m^{(r)} - \boldsymbol{y}_m^{obs} \right\|^2, \quad r = 1, 2, \ldots, N_{bank} \tag{21}$$

where $\boldsymbol{y}_m^{obs}$ is the vector of mean observed or target values of the $m^{th}$ output. Out of all $N_{bank}$ samples, the $P$ samples with the smallest misfit values are selected as

$$\mathcal{P} = \underset{P}{\operatorname{argmin}} \ \left\{ \mathcal{E}^{(r)} \right\}_{r=1}^{N_{bank}} \tag{22}$$

where $\mathcal{P}$ denotes the indices corresponding to the $P$ lowest misfit values. The means and standard deviations of these $P$ samples are then used to initialize the corresponding variational parameters $\boldsymbol{\mu}$ and $\boldsymbol{\sigma}$, respectively. This warm-start strategy reuses information already generated during the screening stage and initializes VI near regions of the reduced input space that produce model outputs close to the available experimental or market targets. In addition, the Sobol-sampled model simulation with the smallest output misfit is defined as the base case

$$r_{base} = \underset{r}{\operatorname{argmin}} \ \mathcal{E}^{(r)}, \qquad \boldsymbol{\theta}_{base} = \boldsymbol{\theta}^{(r_{base})} \tag{23}$$

where $\boldsymbol{\theta}_{base}$ denotes the complete input vector corresponding to the model simulation whose predicted outputs are closest to the experimental or observed data $\boldsymbol{y}^{obs}$.

During the final posterior predictive uncertainty propagation step, the Sobol-screened variables and parameters are sampled from the optimized variational posterior, while the non-selected variables and parameters are fixed at their corresponding base-case values $\boldsymbol{\theta}_{base}$. This allows the posterior uncertainty to be propagated through the high-fidelity model while maintaining a consistent and target-informed reference point for the variables and parameters that were not included in the reduced VI space. For the specific implementation used in this work, the likelihood, prior, and variational distributions are assumed to be Gaussian. Let $K$ denote the number of samples used to estimate the stochastic VI objective. Therefore, the samples from the variational distribution are obtained using the reparameterization, defined as

$$\boldsymbol{\theta}_{N_{top}}^{(k)} = \boldsymbol{\mu} + \boldsymbol{\sigma} \odot \boldsymbol{z}^{(k)}, \qquad \boldsymbol{z}^{(k)} \sim \mathcal{N}(\boldsymbol{0}, \boldsymbol{I}), \qquad k = 1, \dots, K \tag{24}$$

where $\odot$ denotes element-wise multiplication. Subsequently, the corresponding model predictions for the $k^{th}$ variational sample can be represented as:

$$\hat{\boldsymbol{y}}^{(k)} = f\left(\boldsymbol{\theta}_{N_{top}}^{(k)}\right) \tag{25}$$

where $f(\cdot)$ represents the process model or its differentiable surrogate during parameter estimation through VI. Under these assumptions, the simplified negative-ELBO objective function(Zhang et al., 2019) as defined in Eq. (20), after removal of all terms independent of $\boldsymbol{\phi}$, can be expressed as

$$J(\boldsymbol{\mu}, \boldsymbol{\sigma}) = \frac{1}{K} \sum_{k=1}^{K} \sum_{m=1}^{M} \left\| \hat{\boldsymbol{y}}_m\left(\boldsymbol{\theta}_{N_{top}}^{(k)}\right) - \boldsymbol{y}_m^{obs} \right\|^2 + \sum_{j=1}^{N_{top}} \frac{\mu_j^2 + \sigma_j^2}{2\alpha_j^2} - \sum_{j=1}^{N_{top}} \log\ \sigma_j \tag{26}$$

where $\alpha_j$ represents the prior standard deviation of the $j^{th}$ Sobol-selected variable or parameter. Thus, minimizing $J(\boldsymbol{\mu}, \boldsymbol{\sigma})$ simultaneously improves model fit against target data, regularizes the inferred variables and parameters, and promotes a non-degenerate posterior uncertainty representation.

The optimization of the VI objective is carried out using stochastic gradient descent (SGD) based optimization. When the underlying process model is differentiable and derivative information can be accessed directly, AD or solver-provided derivative information may be used to compute gradients of the VI objective with respect to the variational parameters. However, many process simulators, techno-economic models, and external software platforms do not provide gradients in a form directly compatible with modern AD libraries. In such cases, a differentiable surrogate model can be trained using process-model evaluations over the Sobol-selected reduced input space. The trained surrogate provides a differentiable approximation of the input-output mapping over the reduced space, enabling JAX-based AD (Alves et al., 2025; Lin, 2024) during VI. After the surrogate model is trained, the VI objective is minimized using Adam over the variational parameters $\boldsymbol{\phi} = [\boldsymbol{\mu}, \boldsymbol{\sigma}]$. At each iteration, samples are drawn from the standard normal distribution, transformed into samples of the Sobol-selected variables and parameters through the reparameterization, passed through the differentiable surrogate, and used to estimate the stochastic VI loss. Gradients of this loss with respect to the variational parameters are computed using automatic differentiation and used to update $\boldsymbol{\phi}$.

The proposed approach eventually yields the optimized variational parameters $\boldsymbol{\phi}^* = [\boldsymbol{\mu}^*,\ \boldsymbol{\sigma}^*]$ which define the approximate posterior distribution over the selected variables and parameters, given by

$$q\left(\boldsymbol{\theta}_{N_{top}} \mid \boldsymbol{\phi}^*\right) \approx p\left(\boldsymbol{\theta}_{N_{top}} \mid \boldsymbol{\mathcal{D}}\right) \tag{27}$$

Finally, posterior samples are generated from the optimized variational distribution and propagated through the high-fidelity model to evaluate posterior predictive uncertainty. During this final propagation step, the Sobol-selected variables and parameters are sampled from $q\left(\boldsymbol{\theta}_{N_{top}} \mid \boldsymbol{\phi}^*\right)$, while the non-selected variables and parameters are fixed at their corresponding base-case values, as shown in Eq. (23). The resulting predictive distribution can be expressed as $p\left(\hat{\boldsymbol{y}} \mid \boldsymbol{\theta}_{N_{top}}, \boldsymbol{\mathcal{D}}\right)$, where $\hat{\boldsymbol{y}}$ denotes the model predicted process and/or economic outputs. Although a differentiable surrogate is used to enable scalable VI during posterior estimation, final posterior predictive uncertainty is evaluated using the high-fidelity model. This allows VIBES to exploit the fidelity of the underlying process simulator for UQ in process and economic performance metrics, while retaining the scalability (Mukherjee and Zavala, 2026) of surrogate assisted VI.

### 2.3. Multi-Software Integration Platform

The VIBES methodology is independent of a specific process modeling or computational environment. When the process and/or economic model, sensitivity analysis, and Bayesian UQ tools are all available within a common software ecosystem, the framework can be implemented without constructing a multi-software computational platform. For example, a process model developed directly in Python (e.g., using BioSTEAM (Cortes-Peña et al., 2020)) can be coupled directly with scientific computing tools for data analytics, ML, and Bayesian estimation. In many practical process systems engineering applications, however, different software platforms are better suited for different components of the computational workflow. It may be more convenient to undertake rigorous process simulation, flowsheet development, thermodynamic modeling, physical property estimation, and equipment-level calculations in commercial process simulation software like Aspen Plus, Aspen Custom Modeler, gPROMS, Chemcad, AVEVA, etc. Likewise, TEA may be performed using specialized tools such as Aspen Process Economic Analyzer (APEA) or other economic-analysis platforms. In contrast, scientific computing environments such as Python provide flexible access to modern packages for GSA, ML, differentiable programming, stochastic optimization, Bayesian estimation, and visualization.

Selection of software platform(s) can lead to challenges for scalable end-to-end uncertainty-aware process systems engineering tools. From the perspective of Bayesian UQ, it is necessary to have derivative information available from the external process simulator, if used for process/economic modeling, in a form compatible with AD tools. However, for many leading process simulation software, this may be challenging to obtain the desired information in the desired form. Overall, an integration platform is desired to facilitate automated data transfer, repeated model evaluation, output extraction, surrogate model development when required, and posterior predictive uncertainty propagation. VIBES facilitates a multi-software integration platform that can accomplish these desired objectives.

In the present work, the proposed platform is implemented through a Windows Component Object Model (COM) interface using the ‘`win32com`’ package, enabling communication between Aspen Plus (v14.0) and Python (3.12.5). Although several studies have used COM-based interfaces to connect the Aspen ecosystem with Python, either through Aspen tree nodes (Nikkhah et al., 2024) or MS Excel add-ins (Urm et al., 2023), these approaches have generally been limited to separate sampling and optimization workflows. In contrast, this work implements a multi-software integration platform that automates techno-economic model simulations within an end-to-end framework for simultaneous parameter estimation and Bayesian UQ. Specifically, the plant-wide process model for manufacturing KL-SPI bioadhesive is developed in Aspen Plus. Although the corresponding economic analysis can be performed using APEA, for computational flexibility, the economic evaluation is also conducted within the Aspen Plus environment using a cost model implemented through Fortran codes within *Calculator* blocks (Das and Bhattacharyya, 2026b; Turton et al., 2018). This design choice enables process simulation and economic evaluation to be performed within the same executable process modeling environment (i.e., Aspen Plus), thereby allowing automated extraction of both process and economic outputs through a multi-software integration platform represented as the Python-Aspen interface in this paper. The subsequent Sobol GSA, surrogate-assisted VI, posterior estimation, and uncertainty propagation are performed in Python. Thus, the Python-Aspen interface serves as the central computational layer for automated sampling, process simulation, economic model evaluation, sensitivity analysis, Bayesian estimation, and posterior predictive UQ in this work.

The integrated VIBES framework developed in this work is shown in Fig. 1.

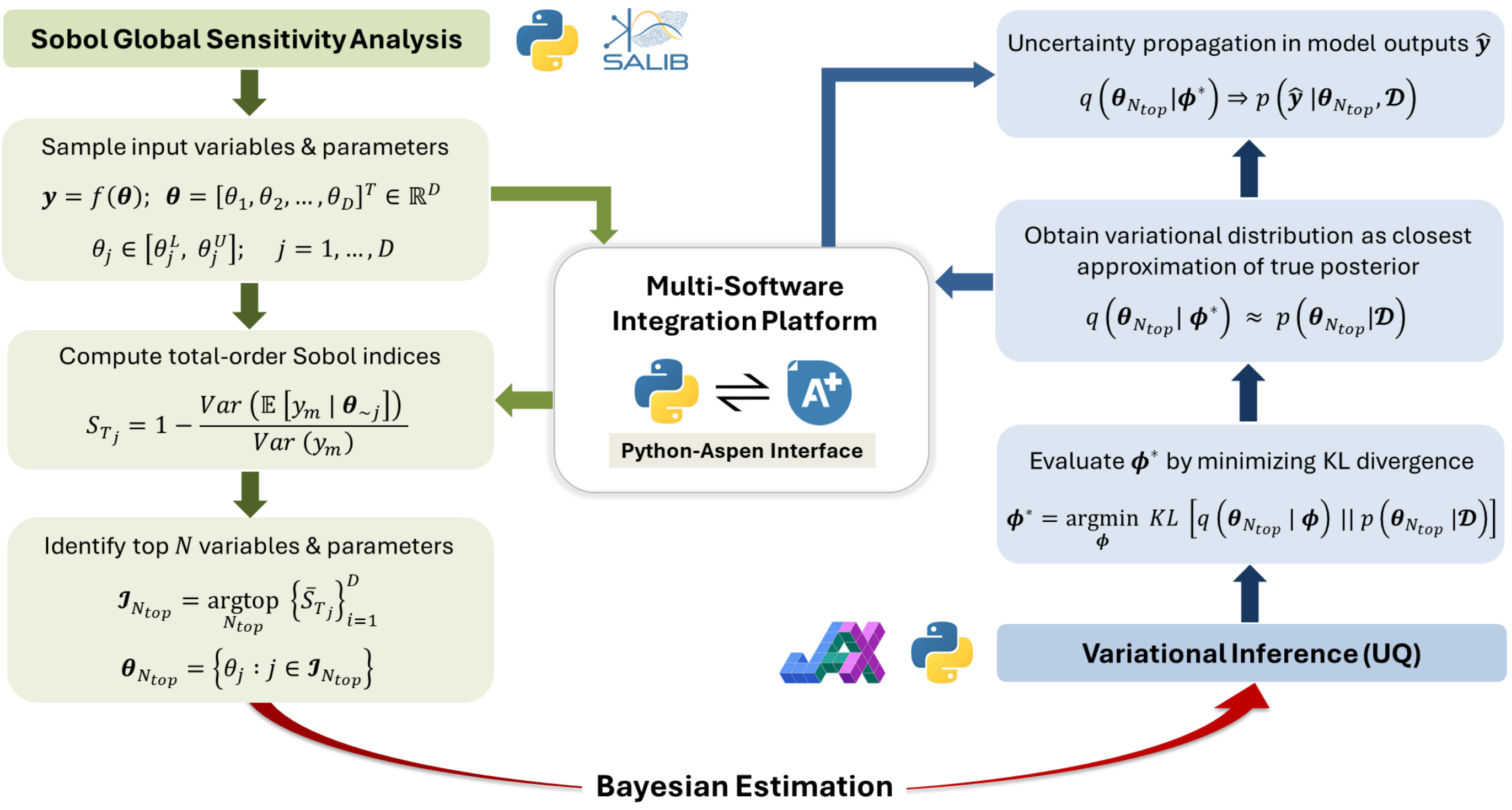

**Fig. 1.** Overall computational workflow for the proposed VIBES algorithm.

## 3. Case Study: Bioadhesive Production Process

The proposed VIBES framework is demonstrated on a lignocellulosic biomass derived formaldehyde-free adhesive production process involving BCD of kraft lignin followed by crosslinking with SPI (Das and Bhattacharyya, 2026; Jiang et al., 2023). The steady-state plant-wide process model is developed in Aspen Plus (v14.0), and a simplified process flow diagram is presented in Fig. 2. The extensive plant-wide modeling and techno-economic optimization framework for the KL-SPI bioadhesive production process under consideration has been discussed by Das and Bhattacharyya (Das and Bhattacharyya, 2026), while the relevant experimental details can be found in the study reported by Jiang et al. (Jiang et al., 2023).

In this process, kraft lignin is first mixed with 0.37 wt.% NaOH (aq) solution at 25 °C and atmospheric pressure in M-101. The resulting mixture is then pumped to 18 atm to facilitate BCD of kraft lignin at 140 to 200 °C while maintaining solvent (i.e., water) in the liquid phase. The pressurized feed is heated to the desired reaction temperature in a shell-and-tube heat exchanger (E-101) using high pressure (HP) steam before entering the depolymerization reactor (R-101). The reactor outlet, containing D-lignin, monomers, small organic compounds (SOCs), and repolymerized species, is cooled to 55 °C in E-102 using cooling water (CW) and sent to a flash drum (D-101), where gaseous by-products are separated from the liquid stream. On the other hand, SPI is dissolved in water at 25 °C and ambient pressure in M-102 and heated to 55 °C in E-103 using low pressure (LP) steam. The D-lignin-rich stream from D-101 is then mixed with the heated SPI solution in an agitated vessel (R-102) to facilitate crosslinking reactions and form the KL-SPI bioadhesive. Excess water is removed from the bioadhesive mixture through controlled heating and flash evaporation in D-102, reducing the solution volume by 75%. Finally, the product stream is cooled to 25 °C in E-107 using CW.

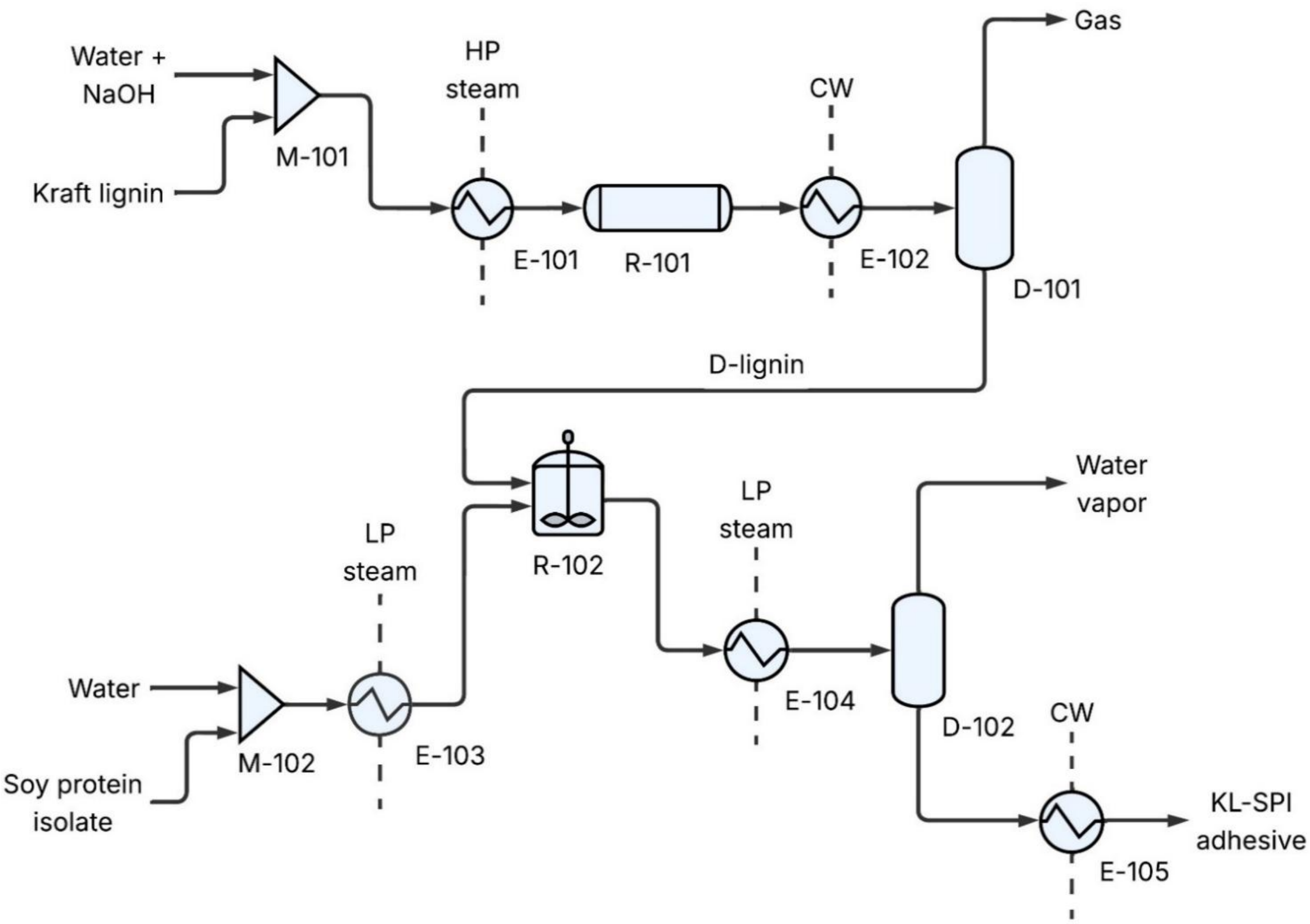

**Fig. 2.** Simplified process flow diagram for KL-SPI bioadhesive manufacturing process considered for implementation of the VIBES framework.

As discussed before, for computational flexibility, the economic evaluation is also implemented within the Aspen Plus environment using a simplified cost model that can generate results consistent with conventional APEA estimates (Das and Bhattacharyya, 2026b; Mevawala et al., 2019; Ogunniyan et al., 2024). The empirical correlations, equipment-specific correction factors, cost equations, and inflation factors required to calculate capital and operating costs are incorporated directly in Aspen Plus using Fortran codes within Calculator blocks (Das and Bhattacharyya, 2026b; Turton et al., 2018). This implementation enables process simulation and economic evaluation to be performed within the same Aspen workflow, thereby facilitating automated output extraction and downstream uncertainty analysis through the Python-Aspen interface. Further details of this economic evaluation framework are discussed extensively in the work by Das and Bhattacharyya (Das and Bhattacharyya, 2026b).

In this work, the proposed VIBES framework requires repeated simulation of the bioadhesive manufacturing process (shown in Fig. 2) in Aspen Plus under different combinations of uncertain kinetic, design/operating, and economic variables and parameters. Therefore, the multi-software integration platform, as discussed in Section 2.3, becomes significantly crucial for integration and automation of Aspen Plus simulations with Python-based sensitivity analysis and Bayesian inference. Thus, the Python-Aspen interface forms the computational backbone of VIBES. As noted earlier, the Python-Aspen interface is implemented via the Windows COM framework. First, the Aspen Plus variable paths to the uncertain inputs and outputs are identified from the Aspen variable explorer. These paths, in turn, allow the workflow to assign sampled inputs, run Aspen simulations, extract model outputs, and save successful simulated results for downstream sensitivity analyses and Bayesian estimation. For example, Code Snippet 1 shows the basic helper functions for establishing the Python-Aspen interaction, such as, accessing Aspen tree nodes (`get_node`), initializing the Aspen Plus archive file from Python (`open_aspen`), and closing the Aspen session (`close_aspen`) after simulation.

Since large-scale sampling analyses may include several input combinations that lead to non-converged Aspen simulations, additional functions are defined to reinitialize or restart Aspen during the automated workflow. As shown in Code Snippet 2, Aspen Plus is reinitialized after each run. The interface first attempts to reinitialize Aspen using available COM methods. However, if reinitialization is insufficient after repeated simulation failures, the Aspen session is closed and restarted from the original archive file. This restart mechanism is helpful to the automated sampling algorithm. After each Aspen simulation, the required process and economic outputs, such as D-lignin mass flow rate and LCOA, are extracted from the corresponding Aspen tree paths and stored for subsequent analysis.

**Code Snippet 1**: Helper functions for developing the Python-Aspen interaction framework.

```
def get_node(aspen, path: str):
    node = aspen.Tree.FindNode(path)
    if node is None:
        raise ValueError(f"Could not find Aspen node: {path}")
    return node
def open_aspen(aspen_file: str):
    aspen = win32.Dispatch("Apwn.Document")
    aspen.InitFromArchive2(os.path.abspath(aspen_file))
    return aspen
def close_aspen(aspen) -> None:
    try:
        aspen.Close()
    except Exception:
        pass
```

**Code Snippet 2**: Reinitialization and restart functions defined for automated Aspen simulations.

```
def reset_aspen_simulation(aspen) -> None:
    try:
        aspen.Reinit()
        return
    except Exception:
        pass
    try:
        aspen.Engine.Reinit()
        return
    except Exception:
        pass
    try:
        aspen.Engine.Reset()
        return
    except Exception:
        pass
def restart_aspen(aspen, aspen_file: str):
    close_aspen(aspen)
    return open_aspen(aspen_file)
```

The proposed approach then proceeds iteratively by passing each sampled input combination to Aspen Plus, followed by simulating the model, extracting outputs, and saving successful runs, while failed simulations are logged separately for diagnostics. The automated simulation approach implemented through the Python-Aspen interface is summarized in Algorithm 1.

The proposed algorithm provides sufficient flexibility to accommodate alternate variance-based sampling methods other than Sobol GSA. Although, the overall computational workflow is demonstrated in Section 2.3 using Sobol GSA (Saltelli et al., 2019, 2010) as the screening step, the proposed approach holds true only when all Aspen Plus generated samples converge. However, if failed Aspen evaluations prevent completion of the structured Sobol design, alternative variance-based screening methods such as the random balance design Fourier amplitude sensitivity test (Goffart and Woloszyn, 2021) (RBD-FAST) can be utilized.

In the present work, the Sobol GSA stage, a total of 20 uncertain input variables and parameters are initially considered, ranging across kinetic, design/operating, and economic sources of uncertainty, as summarized in Table 1. These include feed flow rates, reactor operating conditions, activation energies for the BCD reaction network, raw material prices, utility prices, and other economic parameters. However, among these 20 quantities, three process variables are not treated as independent Sobol variables because they are constrained through design assumptions to ensure process convergence and practical operation, as given by

$$\dot{F}_{SPI} = 2 * \dot{F}_{KL} \tag{28a}$$

$$\dot{F}_{W1} = 20 * \dot{F}_{KL} \tag{28b}$$

$$\dot{F}_{W2} = 10 * \dot{F}_{SPI} \tag{28c}$$

where $\dot{F}_{KL}$, $\dot{F}_{SPI}$, $\dot{F}_{W1}$, and $\dot{F}_{W2}$ represents the mass flow rates of kraft lignin, SPI, kraft lignin solvent (W1), and SPI solvent (W2), respectively. As a result, the effective number of independent uncertain variables and parameters used for Sobol screening becomes $D = 17$. Therefore, for $N = 128$ quasi-random Sobol samples per variable in each base matrix, the total number of Aspen Plus model evaluations ($N_{sim}$) required for this Sobol GSA stage can be calculated as

$$N_{sim} = N * (D + 2) = 128 * (17 + 2) = 2432 \tag{29}$$

**Algorithm 1**: Automated Python-Aspen simulation workflow for variance-based sampling.

---

**Given**: Aspen Plus .bkp file, uncertain input bounds, Aspen variable paths, total number of samples $N_{sam}$, and maximum allowable consecutive convergence failures $N_{fail,max}$

Define the uncertain kinetic, design / operating, and economic variables and parameters

Generate input samples using a variance-based sampling method, e.g., Sobol, RBD-FAST, etc.

Access the Aspen Plus model through the Python-Aspen COM interface

**Initialize**: $n_{sam} = 1, n_{fail} = 0$

**while** $n_{sam} \leq N_{sam}$ **do**

  Select the $n_{sam}$ -th sampled input combination

  Assign sampled input values to the corresponding Aspen Plus input nodes

  Run the Aspen Plus simulation

  **if** Aspen simulation converges successfully **then**

    Extract process and economic outputs from the Aspen Plus output nodes
    Store sampled inputs and extracted outputs as a successful simulation run
    Reinitialize Aspen Plus
    $n_{fail} = 0$

  **else**

    Store sampled inputs and failure message in the failed simulation log
    Reinitialize Aspen Plus
    $n_{fail} = n_{fail} + 1$
    **if** $n_{fail} \geq N_{fail,max}$ **then**

      Close the current Aspen Plus session and restart Aspen Plus from original .bkp file
      $n_{fail} = 0$

    **end if**

  **end if**

  $n_{sam} = n_{sam} + 1$

**end while**

Save the successful simulation dataset and failed simulation log

Close the Aspen Plus session

**Return**: successful simulation dataset (model inputs and outputs), failed simulation log

---

Ideally, Bayesian estimation requires measured experimental data or market-based target data for the process and economic outputs of interest. However, the KL-SPI bioadhesive process considered in this work is an emerging biomass valorization technology, for which practical implementation at pilot or commercial scale remains limited (Das and Bhattacharyya, 2026b).

**Table 1**

List of uncertain input variables and parameters considered for Sobol GSA.

| No. | Input Variables / Parameters | Range | Unit |
|---|---|---|---|
| 1 | Kraft lignin mass flow rate ($F_{KL}$) | [400, 600] | kg/h |
| 2 | W1 / Kraft lignin solvent mass flow rate ($F_{W1}$) | constrained (Eq. 28b) | kg/h |
| 3 | SPI mass flow rate ($F_{SPI}$) | constrained (Eq. 28a) | kg/h |
| 4 | W2 / SPI solvent mass flow rate ($F_{W2}$) | constrained (Eq. 28c) | kg/h |
| 5 | BCD reactor pressure ($P_r$) | [15, 25] | atm |
| 6 | BCD reaction 1 activation energy $\left(E_{a,R1}\right)$ | [30, 70] | kJ/mol |
| 7 | BCD reaction 2 activation energy $\left(E_{a,R2}\right)$ | [25, 45] | kJ/mol |
| 8 | BCD reaction 3 activation energy $\left(E_{a,R3}\right)$ | [25, 45] | kJ/mol |
| 9 | BCD reaction 4 activation energy $\left(E_{a,R4}\right)$ | [25, 45] | kJ/mol |
| 10 | BCD reaction 5 activation energy $\left(E_{a,R5}\right)$ | [25, 45] | kJ/mol |
| 11 | BCD feed temperature ($T_{KL}$) | [140, 200] | ° C |
| 12 | SPI temperature ($T_{SPI}$) | [50, 60] | ° C |
| 13 | Kraft lignin price ($Cost_{KL}$) | [0.15, 0.75] | $/kg |
| 14 | SPI price ($Cost_{SPI}$) | [1.5, 3.0] | $/kg |
| 15 | $NaOH$ price ($Cost_{NaOH}$) | [0.20, 0.65] | $/kg |
| 16 | process water price ($Cost_{W}$) | [1e-4, 3e-4] | $/kg |
| 17 | high pressure (HP) steam price ($Cost_{HPS}$) | [5.0, 12.0] | $/GJ |
| 18 | low pressure (LP) steam price ($Cost_{LPS}$) | [4.0, 7.0] | $/GJ |
| 19 | cooling water (CW) price ($Cost_{CW}$) | [0.1, 0.5] | $/GJ |
| 20 | electricity price ($Cost_{E}$) | [0.05, 0.10] | $/kWh |

Although a few studies have reported the production and evaluation of lignocellulosic biomass-based adhesives (Chen et al., 2024, 2021; Yang and Rosentrater, 2019a, 2019b), available data are not necessarily specific to the KL-SPI bioadhesive production pathway considered here, nor do they provide sufficient process-level and economic target data for the outputs used in this study.

In summary, to the best of our knowledge, there is a very limited availability of experimental and/or market data for directly specifying target values or distributions for D-lignin mass flow rate and LCOA. Therefore, in this work, two data-generation approaches are considered to demonstrate the VI-based inverse UQ workflow. In the first approach, the measured data are represented as single-value target outputs for both D-lignin mass flow rate and LCOA, obtained from the high-fidelity process model and techno-economic optimization framework reported by Das and Bhattacharyya (Das and Bhattacharyya, 2026b) for this specific novel technology, with additional Gaussian noise imposed to represent measurement and market uncertainty. In the second approach, a synthetic-data generation strategy is adopted, where multiple observations are generated to mimic experimental or market data for use during optimization. These synthetic data serve as relevant target data for demonstrating the inverse UQ workflow and can be directly replaced by measured experimental data, pilot-plant data, or market-derived targets when this information becomes available. Specifically, the second strategy involving generation of synthetic measured data is based on propagating uncertainty in the kinetic, design/operating, and economic input variables and parameters through the Aspen Plus process model.

First, feasible ranges are assigned to uncertain inputs based on available literature information for similar systems, raw material and utility price ranges, process operating windows, and kinetic or design assumptions. For instance, the kraft lignin mass flow rate for commercial-scale plants has been reported to be approximately 500–530 kg/h (Ahire et al., 2024; Das and Bhattacharyya, 2026b). Similarly, the ranges for kraft lignin and SPI prices consistent with commercial chemical-market information platforms are approximately 0.25-0.50 $/kg (Abdelaziz et al., 2020; Das and Bhattacharyya, 2026b; Kulas et al., 2021) and 1.8-2.4 $/kg (ChemAnalyst, 2026; Das and Bhattacharyya, 2026b; Elchemy Chemical Market, 2026), respectively. Kinetic uncertainty can be characterized using reported ranges for activation energies of the major reaction steps; for example, published values for lignin depolymerization commonly span approximately 49-55 kJ/mol (Das and Bhattacharyya, 2026b; Forchheim et al., 2014). These ranges for the input variables and parameters consistent with the literature and/or commercial databases provide a basis for defining a plausible design space while generating synthetic data. Representative input samples are then generated from these ranges, for example using Gaussian distributions centered around nominal or literature-informed values, and the corresponding process/economic outputs are obtained from Aspen Plus simulations using the Python-Aspen interface. In this way, synthetic output data are generated systematically from physically meaningful combinations of uncertain input variables and parameters. The underlying premise is that, in the absence of direct experimental or market observations for the output variables, uncertainty in the outputs can be attributed to uncertainty in the underlying input space. Accordingly, the inverse UQ problem seeks to estimate posterior distributions of the dominant variables and parameters that are most consistent with the observed or synthetically generated output targets. This formulation allows the VIBES framework to identify plausible regions of the design and uncertainty space, reduce predictive uncertainty in process and economic metrics, and provide guidance for future experiments or data collection by identifying the variables and parameters that most strongly influence the target outputs.

## 4. Results and Discussions

This section first presents the results of the Sobol GSA used to characterize the uncertainty structure of the bioadhesive production process and identify the dominant set of variables and parameters that govern the target process and economic performance metrics. The resulting reduced input space is subsequently used for VI-based Bayesian inference and posterior predictive UQ, corresponding to both data generation approaches described in Section 3.

### 4.1. Results from Sobol Sensitivity Screening

As discussed in Eq. (29), Sobol GSA is performed over $D = 17$-dimensional independent uncertain input space, thus leading to 2432 Aspen Plus model evaluations generated through the automated Python-Aspen interface. As noted earlier, although Table 1 lists 20 uncertain input variables and parameters, 3 mass flow rates are determined through the design constraints as defined in Eqs. (28a)–(28c) and are therefore not independently sampled during Sobol screening. Overall, Fig. 3 illustrates the sampled values of all 20 input variables and parameters including 17 independent inputs sampled from Sobol GSA, and 3 dependent inputs corresponding to the design constraints under consideration. These samples define the prior uncertainty bounds of the input variables and parameters before Bayesian estimation.

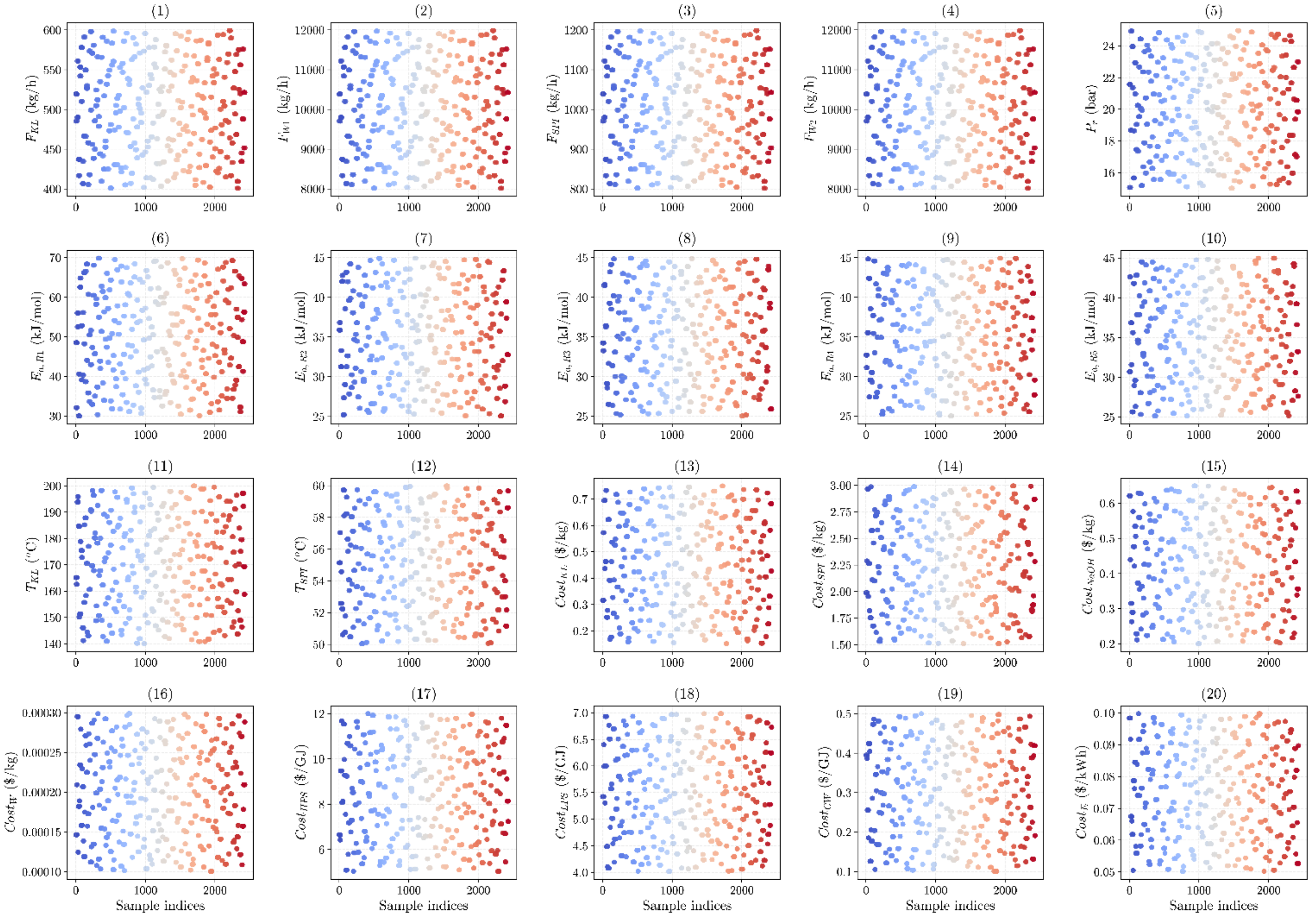

**Fig. 3.** Prior samples of all 20 uncertain input variables and parameters obtained from Sobol GSA, labeled according to the respective numbering defined in Table 1.

The corresponding economic and process output results obtained from the successful Aspen Plus simulations are shown in Fig. 4. These results of the output variables, namely the LCOA (that characterizes economic performance of the KL-SPI production pathway) and D-lignin mass flow rate (that characterizes process performance), are used to generate the prior predictive uncertainty bounds of model outputs before Bayesian inference. Using the successful Sobol samples, the average total-order Sobol indices $\left(\bar{S}_{T_j}\right)$ defined in Eq. (11) are computed with respect to both model outputs.

Since the cumulative contribution of the top 5 highest-ranked input variables and parameters based on $\bar{S}_{T_j}$ captures more than 99% of the total averaged output variance, in this work, the reduced input space dimension is selected as $N_{top} = 5$. Table 2 lists this sequence of 5 dominant variables and parameters ranked according to their respective $\bar{S}_{T_j}$ values. Accordingly, the reduced input vector, as defined in Eq. (13), to be considered for the subsequent VI stage is given by

$$\boldsymbol{\theta}_{N_{top}} = \boldsymbol{\theta}_5 = \left\{F_{KL},\ \ Cost_{SPI},\ \ Cost_{KL},\ \ E_{a,R1},\ \ E_{a,R2}\right\}^T \tag{30}$$

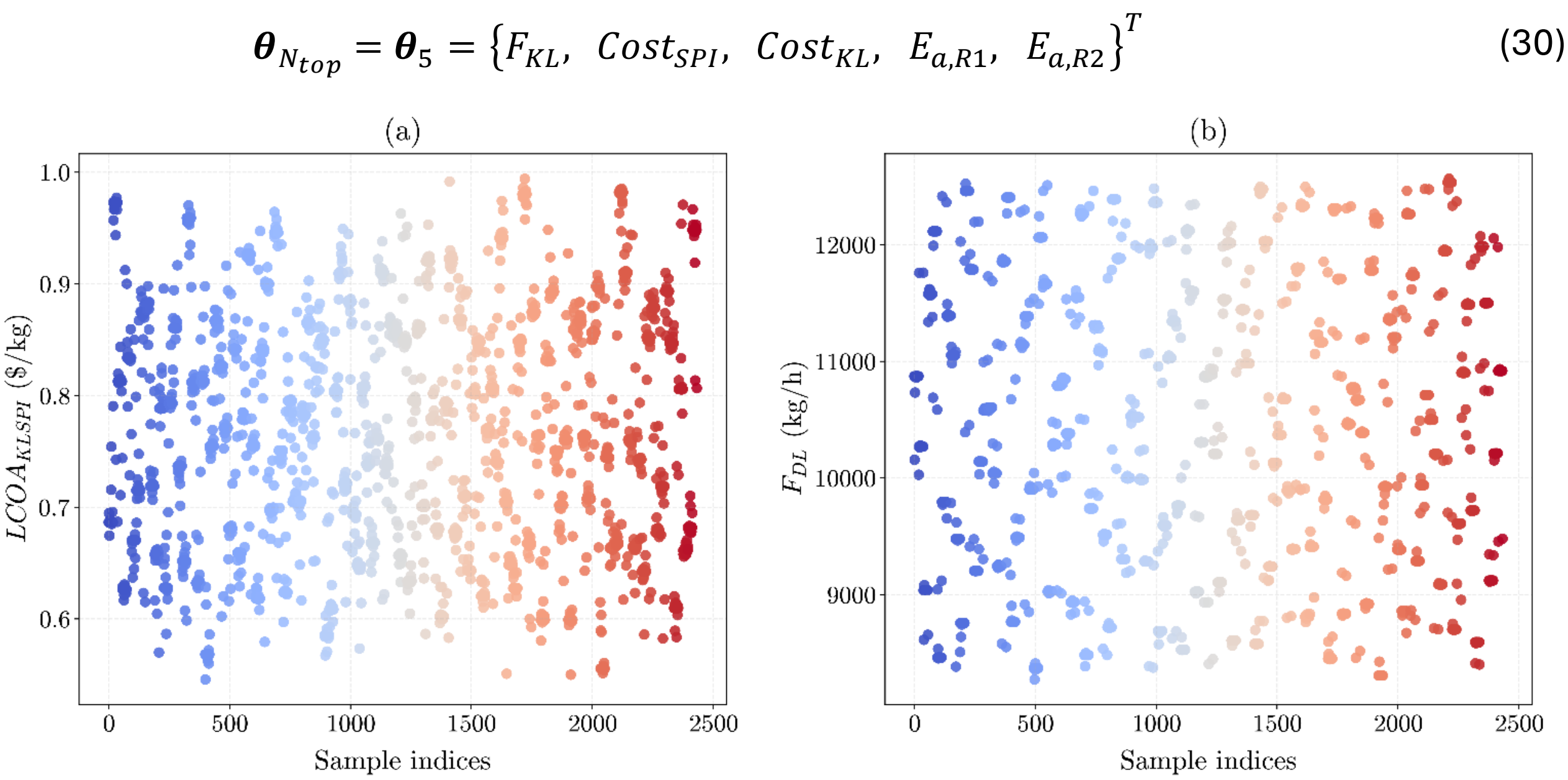

**Fig. 4.** Prior predictive results of model outputs represented by **(a)** KL-SPI bioadhesive levelized cost and **(b)** D-lignin mass flow rate obtained from Sobol GSA.

The results obtained from Sobol GSA confirm that a relatively small subset of uncertain variables and parameters governs the majority of uncertainty in the selected process and economic outputs. This sensitivity-driven reduction avoids performing Bayesian estimation over the full input space and provides a computationally tractable basis for subsequent posterior estimation and predictive uncertainty propagation.

### 4.2. Results from VI

This section presents the VI-based Bayesian estimation and posterior predictive UQ results obtained after Sobol screening. In this work, the surrogate is constructed as a feedforward neural network (NN) model with two hidden layers, eight neurons per hidden layer, and hyperbolic tangent activation functions (Mukherjee and Bhattacharyya, 2023). The input and output data are normalized before training, and the surrogate parameters are optimized using the Adam optimizer.

**Table 2**

List of top 5 dominant uncertain input variables and parameters identified through Sobol GSA, ranked as per their respective averaged total-order Sobol indices.

| Rank | Selected Input Variable / Parameter | $\bar{S}_{T_j}$ |
|---|---|---|
| 1 | KL mass flow rate ($F_{KL}$) | 0.508 |
| 2 | SPI price ($Cost_{SPI}$) | 0.445 |
| 3 | KL price ($Cost_{KL}$) | 0.018 |
| 4 | BCD reaction 1 activation energy $\left(E_{a,R1}\right)$ | 0.014 |
| 5 | BCD reaction 2 activation energy $\left(E_{a,R2}\right)$ | 0.009 |

As discussed in Section 4.1, Sobol GSA identifies the dominant uncertain variables and parameters governing the D-lignin mass flow rate and KL-SPI bioadhesive LCOA. Within the proposed framework, variability in process and economic performance is interpreted as the combined effect of uncertainty in kinetic, design/operating, and economic variables and parameters. Accordingly, the inverse UQ problem seeks posterior distributions of the Sobol-selected inputs that are consistent with the observed ranges of the target model outputs. The warm-start strategy described in Section 2.2 is implemented for both data-generation approaches considered in the VI-based inverse UQ workflow. For the single-value-target approach, the warm-start samples are selected according to their residuals relative to the specified (true) target outputs obtained from literature-reported simulation of the novel high-fidelity process model for the bioadhesive production technology (Das and Bhattacharyya, 2026b) considered in this work. However, for the synthetic-data approach, the base case is identified using the residual with respect to the mean value of the corresponding generated (measurement) outputs mimicking experimental or market data. This base case provides the fixed values of non-selected variables and parameters during final posterior predictive uncertainty propagation through the high-fidelity Aspen Plus model. Additionally, for the second approach (i.e., synthetic data generation), the dominant uncertain variables and parameters are sampled from normal distributions whose means and variances are selected using ranges reported in available literature and public databases (Ahire et al., 2024; ChemAnalyst, 2026; Das and Bhattacharyya, 2026b; Forchheim et al., 2014; Kulas et al., 2021). In this work, a total of 30 synthetic observation samples is used for Bayesian estimation and UQ.

For both the approaches considered in this paper, sensitivity analyses are performed to quantify and contrast the relative uncertainty reductions in both posterior parameter / variable and predictive distributions with respect to the dimensionality of the reduced input space, i.e., $N_{top}$. Corresponding to each value of $N_{top} = 1$, 2, ..., 5, the surrogate NN model described in at the beginning of this section is trained for the Sobol-sampled simulation data for 5000 epochs. The respective training computational (CPU) times remain approximately consistent at 70–80 seconds across the different reduced input space dimensionality values. For instance, the surrogate NN model trained for the reduced input vector defined in Eq. (30), i.e., $\boldsymbol{\theta}_5$ for $N_{top} = 5$, yields root mean squared error (RMSE) values of around 0.009 $\$/\mathrm{kg}$ (approximately 1.15 %) and 27.5 $\mathrm{kg/h}$ (approximately 0.26 %) for LCOA and D-lignin mass flow rate, respectively. These results indicate that the surrogate model provides an accurate approximation of the high-fidelity Aspen Plus model for gradient-based VI optimization. For all cases, VI is performed over the Sobol-screened reduced input space using mini-batch SGD with the Adam optimizer. In this work, 5 variational samples are used per mini-batch, and optimization is conducted for 1000 epochs subject to the same initialization using gradients obtained through JAX-based AD of the trained surrogate model. After convergence, 500 samples are drawn from the learned variational distribution and propagated through the original Aspen Plus model using the Python-Aspen interface to quantify posterior predictive uncertainty. The corresponding results obtained from both the approaches considered in this work are discussed in subsequent sections.

#### 4.2.1. Approach 1: Single-Value Target Data

The first approach considers a single value each for the output economic and process performance metrics. Based on the simulation of the high-fidelity process model developed by Das and Bhattacharyya (Das and Bhattacharyya, 2026b) for the same KL-SPI bioadhesive production technology considered in this work, the target (true) data for the model outputs, namely LCOA and D-lignin mass flow rate, respectively can be represented as

$$y_{true} = [0.785, \ \ 10686.013]^T \tag{31}$$

In addition, a Gaussian measurement noise is added to $y_{true}$ to represent experimental and market uncertainty. For the reduced input space defined in Eq. (30), i.e., $N_{top} = 5$, Fig. 5 compares the prior and posterior predictive distributions of the two output variables.

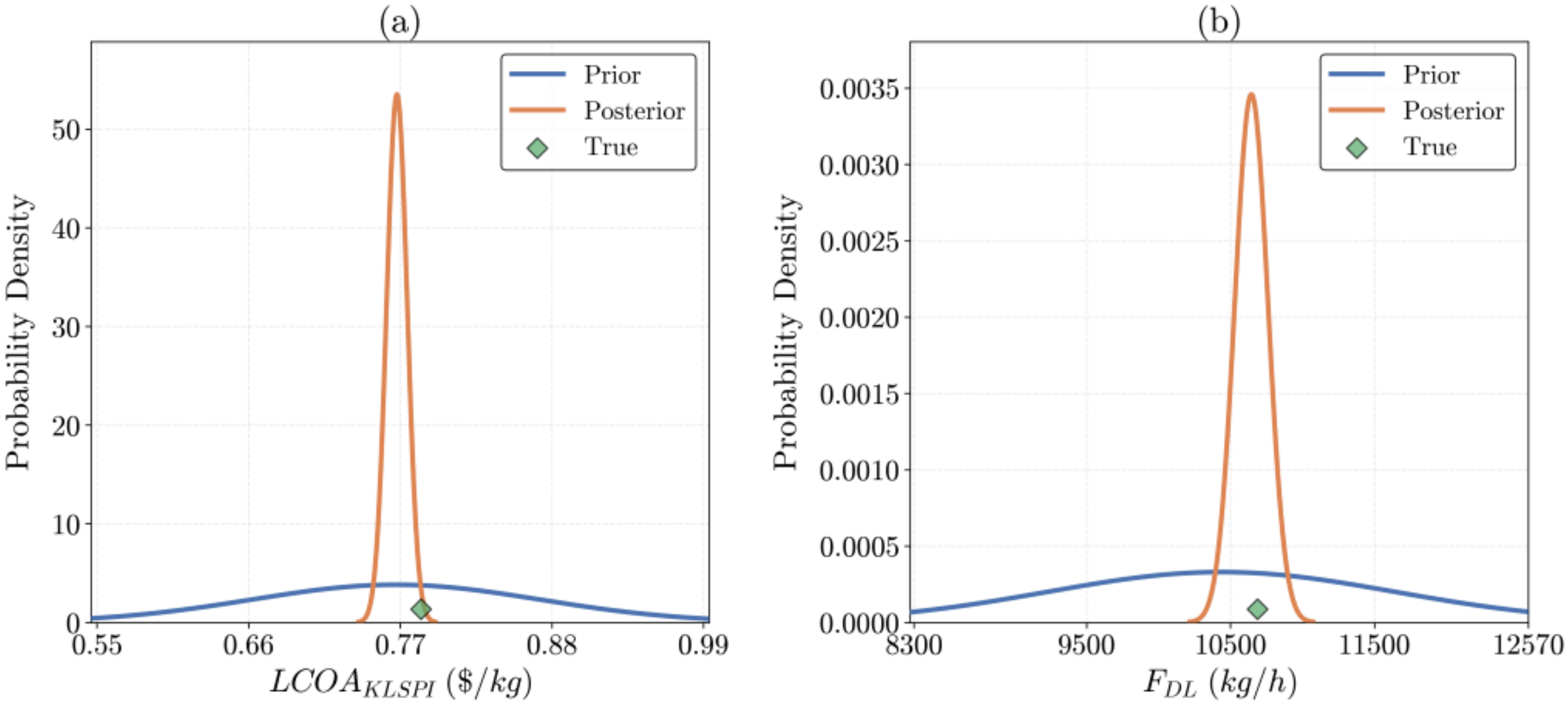

**Fig. 5.** Prior and posterior predictive distributions of (a) economic (i.e., levelized cost of KL-SPI bioadhesive) and (b) process (i.e., D-lignin mass flow rate) outputs for Approach 1 with $N_{top} = 5$. The horizontal limits indicate the range of prior predictive output results obtained from Sobol GSA.

The prior samples are obtained by propagating the Sobol-screened input uncertainty through the Aspen Plus model, whereas the posterior samples are obtained by propagating samples from the optimized variational distributions. The posterior predictive distributions in Figs. 5(a) and 5(b) exhibit substantial contractions relative to the corresponding prior predictive distributions. Specifically, the uncertainty bounds for LCOA, and D-lignin mass flow rate are reduced by approximately 86.3 % and 85.5 %, respectively. These results indicate that Bayesian estimation over the Sobol-selected reduced space can substantially reduce uncertainty in both economic and process output variables while retaining the high-fidelity Aspen Plus model for final uncertainty propagation. Fig. 6 presents the corresponding prior and posterior distributions of the 5 Sobol-selected variables and parameters, as discussed in Eq. (30). These results illustrate how the available target information updates the initial uncertainty characterization and identifies posterior regions of the reduced input space that are consistent with the specified output targets. To evaluate the influence of reduced input-space dimension on posterior contraction, Table 3 summarizes the percentage reduction in uncertainty bounds for the Sobol-selected variables and parameters across $N_{top}$= 1, ..., 5.

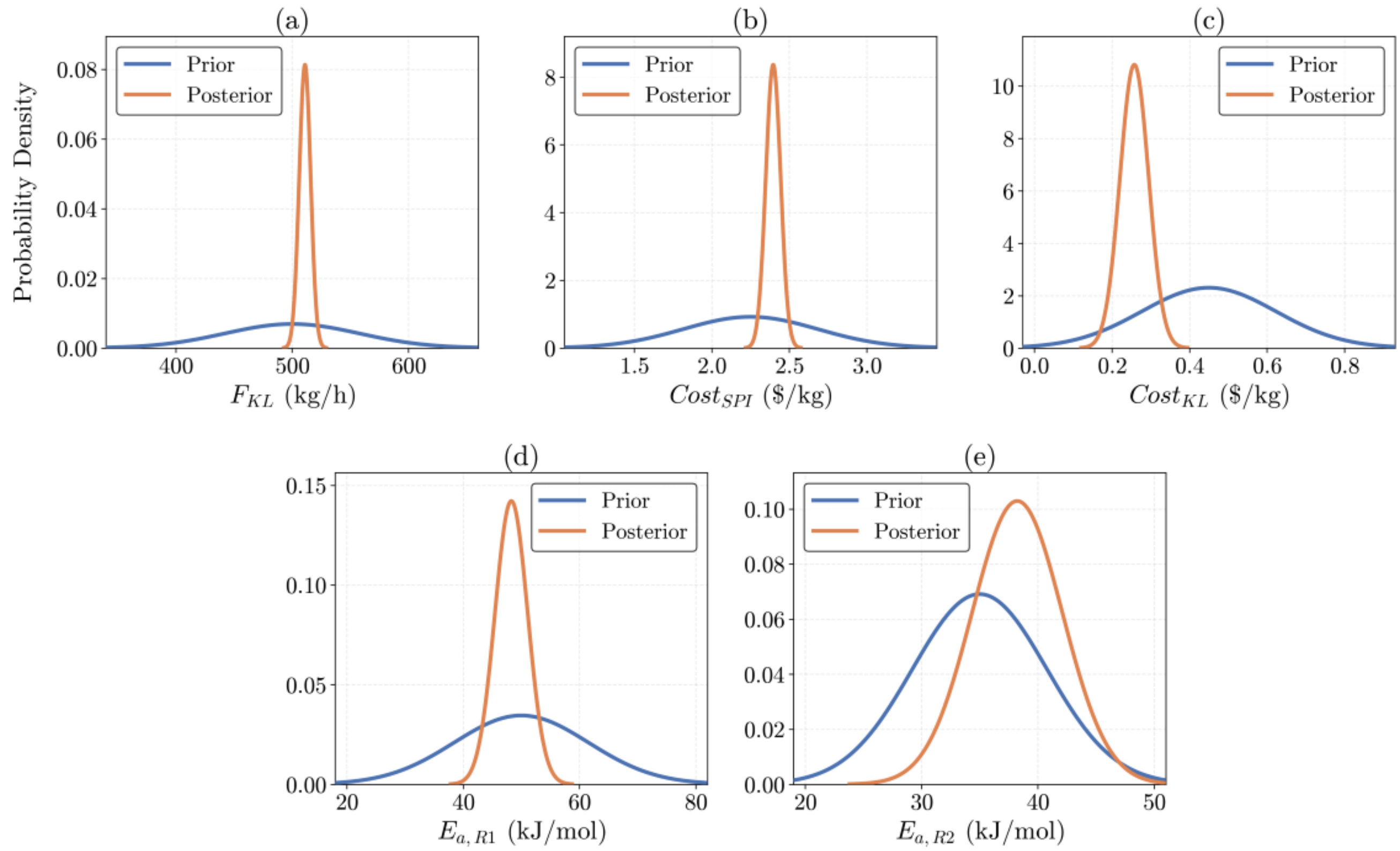

**Fig. 6.** Prior and posterior distributions of **(a)** KL mass flow rate, **(b)** SPI cost, **(c)** KL cost, **(d)** BCD reaction 1 activation energy, and **(e)** BCD reaction 2 activation energy for Approach 1.

As additional variables and parameters are incorporated into the VI space, the dominant variables continue to exhibit appreciable posterior contraction. In contrast, the less influential variables and parameters added at larger $N_{top}$ values generally show comparatively smaller uncertainty reductions, as expected from their lower averaged total-order Sobol indices $\left(\bar{S}_{T_j}\right)$ and limited contributions to the total output variance. The VI loss trajectory for Approach 1 corresponding to $N_{top} = 5$ is presented in Fig. A.1(a) in Appendix A. The comparison among the prior and posterior parameter/variable and predictive results obtained from the proposed framework for this case, along with the respective means and variances, are shown in Figs. A.2 and A.3, in Appendix A.

**Table 3**

Percentage reduction in uncertainty bounds of the Sobol-selected variables and parameters for Approach 1 across different reduced-space dimensions.

| **Variable / Parameter** | $N_{top} = 1$ | $N_{top} = 2$ | $N_{top} = 3$ | $N_{top} = 4$ | $N_{top} = 5$ |
|---|---|---|---|---|---|
| $F_{KL}$ | 88.8 % | 82.4 % | 85.8 % | 85.0 % | 84.6 % |
| $Cost_{SPI}$ | -- | 76.8 % | 81.0 % | 81.6 % | 77.9 % |
| $Cost_{KL}$ | -- | -- | 50.4 % | 68.8 % | 54.2 % |
| $E_{a,R1}$ | -- | -- | -- | 71.9 % | 57.8 % |
| $E_{a,R2}$ | -- | -- | -- | -- | 9.2 % |

Additionally, Table 4 compares the reduction in posterior predictive uncertainty for LCOA and D-lignin mass flow rate across different values of $N_{top}$, together with the corresponding computational (CPU) time associated with VI optimization. The last row of Table 4 reports the results obtained when Bayesian estimation is performed over the full uncertain input space, i.e., $N_{top} = D = 17$.

**Table 4**

Posterior predictive UQ and computational expense comparison for Approach 1 across different reduced-space dimensions.

| **Reduced-space dimension** | **Uncertainty reduction in LCOA of KL-SPI** | **Uncertainty reduction in D-lignin mass flow rate** | **Computational time for VI** |
|---|---|---|---|
| $N_{top} = 1$ | 98.8 % | 88.9 % | 4.8 min |
| $N_{top} = 2$ | 84.2 % | 83.6 % | 4.7 min |
| $N_{top} = 3$ | 90.7 % | 83.1 % | 4.6 min |
| $N_{top} = 4$ | 93.0 % | 85.2 % | 4.7 min |
| $N_{top} = 5$ | 86.3 % | 85.5 % | 4.6 min |
| $N_{top} = D = 17$ | 82.3 % | 85.4 % | 6.7 min |

In summary, these results demonstrate that the Sobol-selected reduced variable and parameter space can achieve substantial computational savings relative to the full uncertain input space while still accurately representing uncertainties in the outputs. In particular, the five-dimensional reduced space, i.e., $N_{top} = 5$, captures more than $99\ \%$ of the average output variance while significantly reducing the computational expense by approximately $31.3\ \%$ relative to the full-space case, i.e., $N_{top} = D = 17$.

For additional comparison, the posterior parameter/variable and predictive distributions for $N_{top} = 2$, which captures approximately 95 % of the average output variance, are provided in Figs. A.4 and A.5 respectively, in Appendix A. It is worth noting that for this case study, selecting $N_{top} = 1$ is not sufficient to adequately represent the reduced input-output relationship, likely because the single highest-ranked screened variable (i.e., $F_{KL}$) accounts for only approximately $51\ \%$ of the average output variance, as evident from Table 2. Moreover, the corresponding optimal surrogate NN model yields RMSE values of around $13.1\ \%$ for LCOA and $0.6\ \%$ for D-lignin mass flow rate. Consequently, although Table 4 indicates significant posterior uncertainty reduction for $N_{top} = 1$, the resulting posterior predictions are not satisfactorily consistent with the target data, as illustrated in Fig. A.6 in Appendix A. This limitation/inadequacy is more pronounced for the synthetic-data generation approach discussed in the next subsection, where the posterior predictions should ideally be capable of reproducing the variability across multiple representative measurements.

#### 4.2.2. Approach 2: Synthetic Measurement Data

The second approach considers multiple synthetic measurements of the economic and process outputs to mimic the availability of experimental or market data. Following simulation of the high-fidelity Aspen model using literature-reported values of the dominant inputs, as discussed in Section 3, additional measurement noise is introduced before optimization. The noise is modeled as zero-mean Gaussian distributions with standard deviations corresponding to 0.5% of the mean simulated LCOA and 1.0% of the mean simulated D-lignin mass flow rate. For the five-dimensional reduced input space defined in Eq. (30), Fig. 7 compares the prior and posterior predictive distributions for LCOA and D-lignin mass flow rate, respectively. The posterior predictive distributions for both the outputs are compared with the distribution of synthetic data used during VI. It is observed that the inferred reduced-space distributions provide output predictions consistent with the available measurement variability.

Similar to Approach 1, the posterior predictive distributions exhibit substantial contraction relative to the prior uncertainty ranges while remaining informed by the synthetic observations. Specifically, for $N_{top} = 5$, the uncertainty bounds of LCOA and D-lignin mass flow rate are reduced by 90.6 % and 93.9 % respectively, as summarized in Table 6. This result indicates that the availability of multiple measurements provides a more informative inverse UQ workflow than the single-value-target case, thus enabling the posterior distributions to be constrained by the observed variability of both process and economic outputs. Fig. 8 presents the corresponding prior and posterior distributions of the five Sobol-selected variables and parameters. The posterior distributions exhibit significant contraction relative to their prior ranges, indicating how such measurements, if available, can provide valuable information to update the dominant kinetic, design/operating, and economic variables and parameters. The VI loss trajectory corresponding to $N_{top} = 5$ for Approach 2 is shown in Fig. A.1(b) in Appendix A. Similar to Approach 1, the comparison among the prior and posterior parameter/variable and predictive results obtained from the proposed VIBES framework for this case, along with the respective means and variances, are shown in Figs. A.7 and A.8, in Appendix A.

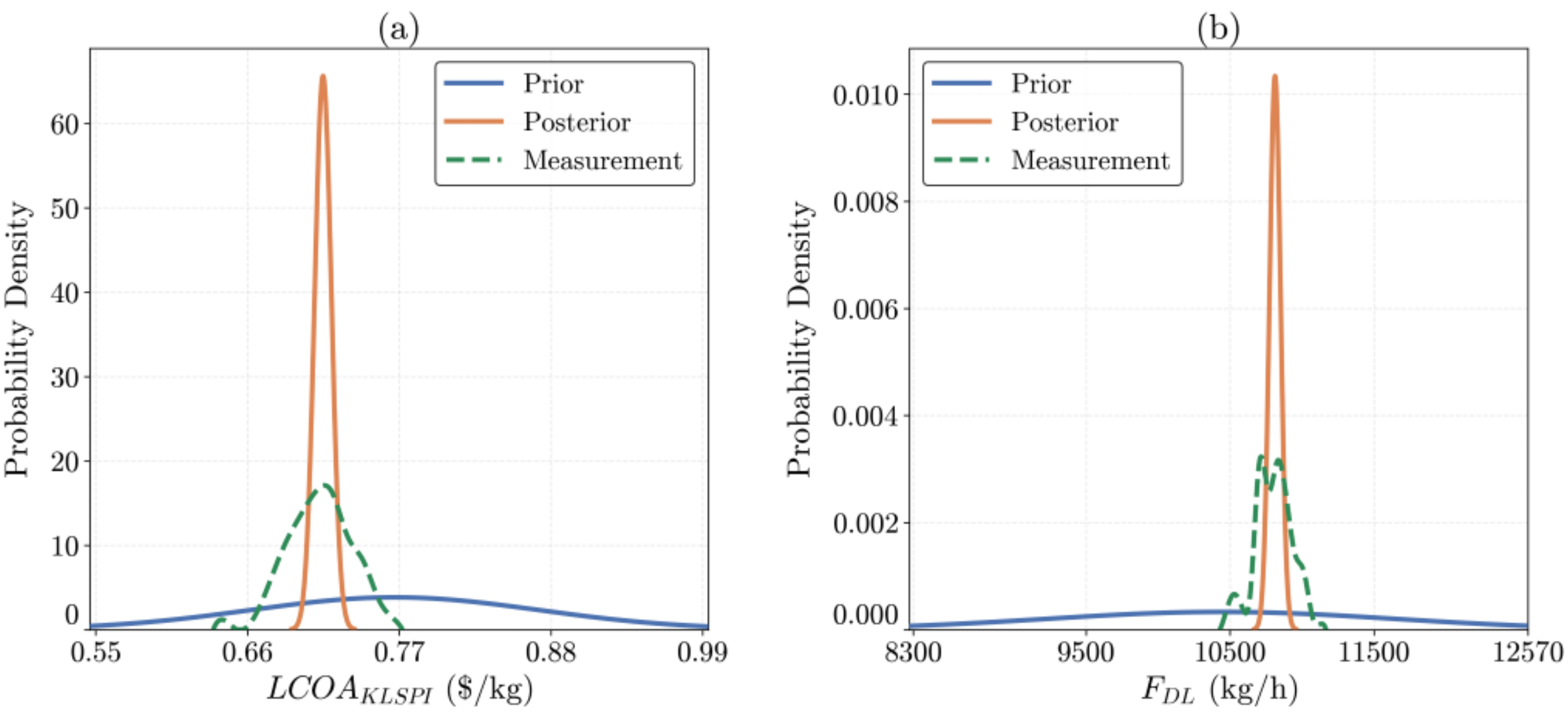

**Fig. 7.** Prior and posterior predictive distributions of (a) economic (i.e., levelized cost of KL-SPI bioadhesive) and (b) process (i.e., D-lignin mass flow rate) outputs, along with the distribution of synthetic measured data used for VI in Approach 2 with $N_{top} = 5$. The horizontal limits indicate the range of prior predictive output results obtained from Sobol GSA.

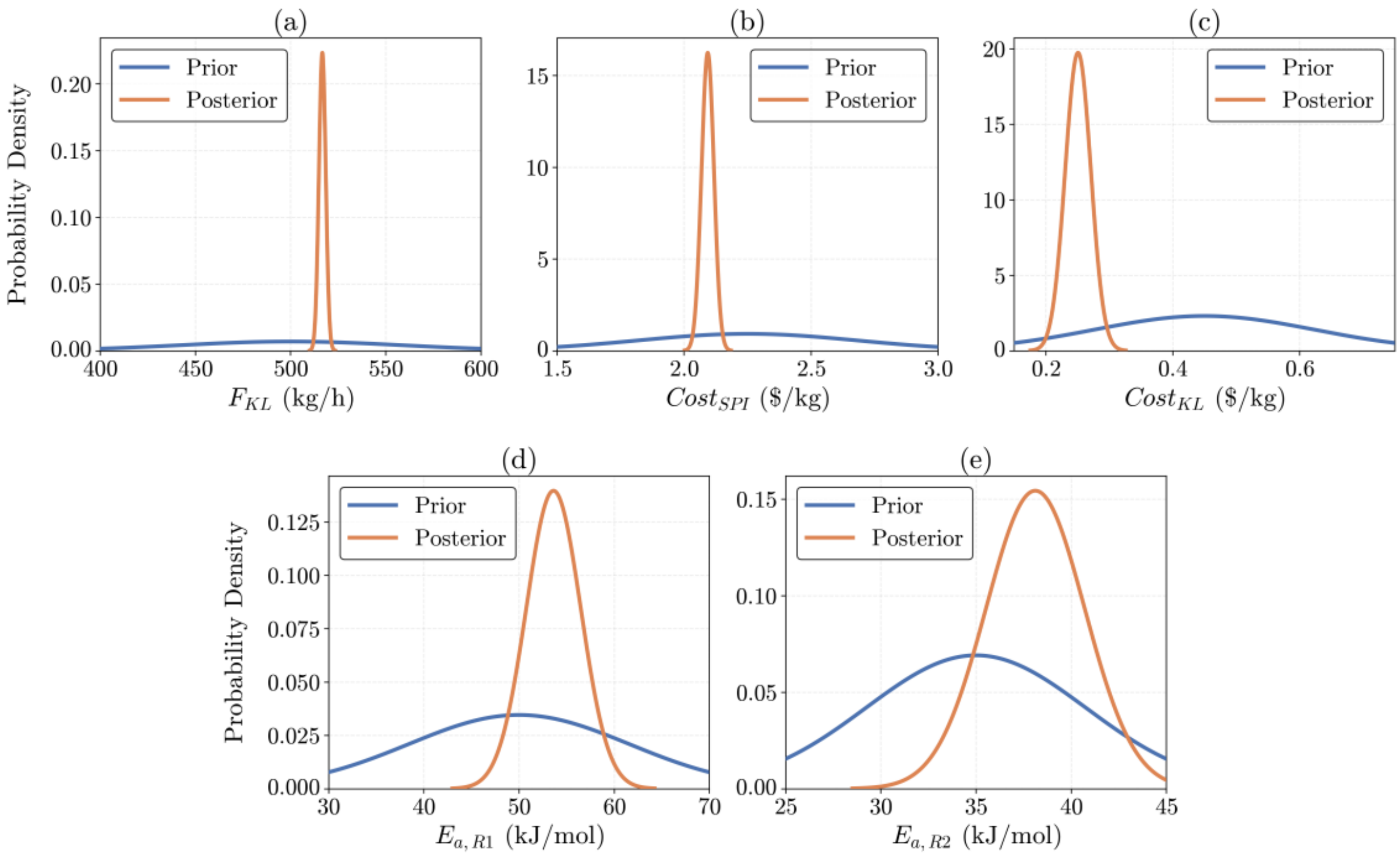

**Fig. 8.** Prior and posterior distributions of **(a)** KL mass flow rate, **(b)** SPI cost, **(c)** KL cost, **(d)** BCD reaction 1 activation energy, and **(e)** BCD reaction 2 activation energy for Approach 2.

Table 5 summarizes the percentage reduction in uncertainty bounds of the Sobol-selected variables and parameters for different reduced-space dimensions $(N_{top})$. Availability of multiple measurements results in a more informed inverse UQ problem, and the posterior uncertainty reductions are observed to be consistently higher than those obtained using the single-value-target formulation in Approach 1. For example, in the $N_{top} = 5$ scenario, the uncertainty bounds of $F_{KL}$, $Cost_{SPI}$, $Cost_{KL}$, $E_{a,R1}$, and $E_{a,R2}$ are reduced by 94.5 %, 90.6 %, 79.7 %, 58.6 %, and 44.6 %, respectively.

Table 6 compares posterior predictive uncertainty reduction and VI computational time across the different values of $N_{top}$. Across all reduced-space dimensionalities under consideration, the uncertainty reduction in LCOA ranges from 85.9 % to 99.2 %, while the corresponding reduction in D-lignin mass flow rate ranges from 91.2 % to 95.6 %. In addition, for the $N_{top} = 5$ case, the VI records an approximate computational (CPU) time of 4.1 min, as compared to 6.3 min for the full-space case with $N_{top} = D = 17$.

**Table 5**

Percentage reduction in uncertainty bounds of the Sobol-selected variables and parameters for Approach 2 across different reduced-space dimensions.

| Variable / Parameter | $N_{top} = 1$ | $N_{top} = 2$ | $N_{top} = 3$ | $N_{top} = 4$ | $N_{top} = 5$ |
|---|---|---|---|---|---|
| $F_{KL}$ | 95.5 % | 93.8 % | 93.7 % | 91.5 % | 94.5 % |
| $Cost_{SPI}$ | -- | 82.8 % | 83.4 % | 79.6 % | 90.6 % |
| $Cost_{KL}$ | -- | -- | 83.7 % | 79.9 % | 79.7 % |
| $E_{a,R1}$ | -- | -- | -- | 73.5 % | 58.6 % |
| $E_{a,R2}$ | -- | -- | -- | -- | 44.6 % |

**Table 6**

Posterior predictive UQ and computational expense comparison for Approach 2 across different reduced-space dimensions.

| Reduced-space dimension | Uncertainty reduction in LCOA of KL-SPI | Uncertainty reduction in D-lignin mass flow rate | Computational time for VI |
|---|---|---|---|
| $N_{top} = 1$ | 99.2 % | 95.6 % | 4.2 min |
| $N_{top} = 2$ | 86.3 % | 94.3 % | 4.0 min |
| $N_{top} = 3$ | 87.9 % | 94.1 % | 3.9 min |
| $N_{top} = 4$ | 85.9 % | 91.2 % | 4.0 min |
| $N_{top} = 5$ | 90.6 % | 93.9 % | 4.1 min |
| $N_{top} = D = 17$ | 85.7 % | 92.0 % | 6.3 min |

Thus, the Sobol-selected five-dimensional space, that captures more than 99 % of the mean output variance, reduces the VI computational expense by approximately 34.9 % relative to the full uncertain input space while retaining substantial posterior predictive uncertainty reduction. For clarity, additional posterior parameter/variable and predictive distributions for $N_{top} = 2$ are also provided in Figs. A.9 and A.10 respectively, in Appendix A. Finally, similar to Approach 1, it is worth noting that the $N_{top} = 1$ case is not sufficient to accurately represent the inverse UQ problem because the highest-ranked variable alone accounts for only approximately 51 % of the mean output variance. Although Table 6 reports substantial posterior uncertainty reduction for this case, the resulting posterior predictions do not adequately accommodate the available synthetic measurement distribution, as illustrated in Fig. A.11 in Appendix A. This result further demonstrates that down-selection of the input space must be appropriately done with due consideration of the resulting output variance.

## 5. Conclusions and Future Work

This work developed VIBES, a two-stage scalable Bayesian framework that integrates Sobol GSA with VI for simultaneous estimation and UQ of uncertain input variables and parameters in high-dimensional process systems. VIBES combines Sobol GSA with VI to first identify a subset of uncertain inputs that most strongly influence target process and economic outputs, followed by estimating their posterior distributions within a scalable reduced dimensional space. By avoiding Bayesian estimation over the full uncertain input space, the framework improves computational tractability while retaining the dominant sources of predictive uncertainty. The learned posterior distributions are subsequently propagated through the high-fidelity process model to quantify posterior predictive uncertainty in process and economic performances.

The proposed framework is demonstrated for a KL-SPI bioadhesive production process, considering D-lignin mass flow rate and LCOA as the process and economic outputs of interest. The framework is implemented through an automated Python-Aspen interface, enabling structured input sampling, high-fidelity process simulation, output extraction, sensitivity screening, Bayesian estimation, and final posterior predictive uncertainty propagation through the Aspen Plus model. Sobol screening of the 17-dimensional independent uncertain input space identified kraft lignin mass flow rate, SPI price, kraft lignin price, and the activation energies of BCD reactions 1 and 2 as the five dominant variables and parameters. Collectively, these quantities captured more than (99%) of the mean output variance, thereby providing a substantially lower-dimensional yet informative space for Bayesian estimation.

Subsequently, VI-based Bayesian estimation is undertaken by using both single-value target data and multiple synthetic measurements due to limited experimental and market information. In both approaches, posterior propagation through the high-fidelity Aspen Plus model produces substantial reductions in predictive uncertainty for LCOA and D-lignin mass flow rate. For the five-dimensional ($N_{top} = 5$) reduced space, the single-value-target approach reduces predictive uncertainty by 86.3 % for LCOA and 85.5 % for D-lignin mass flow rate, while reducing VI computational time by approximately 31.3 % relative to inference over the full ($N_{top} = D$ =17) uncertain input space. When multiple synthetic measurements are considered, the posterior estimates are more strongly informed by the available output variability, yielding uncertainty reductions of 90.6 % for LCOA and 93.9 % for D-lignin mass flow rate, together with an approximately 34.9 % reduction in VI computational time relative to the full-space scenario. As expected, these results demonstrate that multiple measurements can provide a more informative inverse estimation and UQ than single-value targets. The results also highlight that posterior uncertainty contraction alone is not sufficient indicator of satisfactory UQ. In particular, the $N_{top} =$ 1 case exhibits substantial apparent uncertainty reduction but does not adequately represent the available target or measurement data, because the highest-ranked variable alone accounted for approximately 51 % of the mean output variance. Thus, selecting an appropriate reduced-space dimension is essential to balance computational efficiency with the ability to represent the coupled process and economic output variability.

Future work will investigate additional Aspen-Python interface strategies that may reduce or eliminate reliance on differentiable surrogate models during Bayesian estimation. In particular, the use of derivative information available through Aspen EO sensitivity analysis, or related gradient-based optimization capabilities, could enable more direct scalable optimization using the high-fidelity process model. Further work will also extend VIBES to larger and more highly interconnected process flowsheets. Although demonstrated here for KL-SPI bioadhesive production, the proposed framework is generic and readily extendable to other complex process systems. More broadly, VIBES provides a scalable pathway for integrating GSA, Bayesian estimation, and posterior predictive UQ with rigorous process simulation and techno-economic analysis to support uncertainty-aware process design, model calibration, and decision-making.

## Code Availability Statement

All codes and data required to implement the proposed VIBES algorithm can be accessed through a public repository on GitHub, which will be made available upon acceptance of the paper.

## Acknowledgements

The project is funded by the USDA-National Institute of Food & Agriculture through the project titled "Mid-Atlantic Sustainable Biomass for Value-Added Products Consortium (MASBio)" (Grant #2020-68012-31881). The USDA financial support is gratefully acknowledged.

## Appendix

This section contains additional results obtained from the VIBES algorithm proposed in this work.

### Appendix A

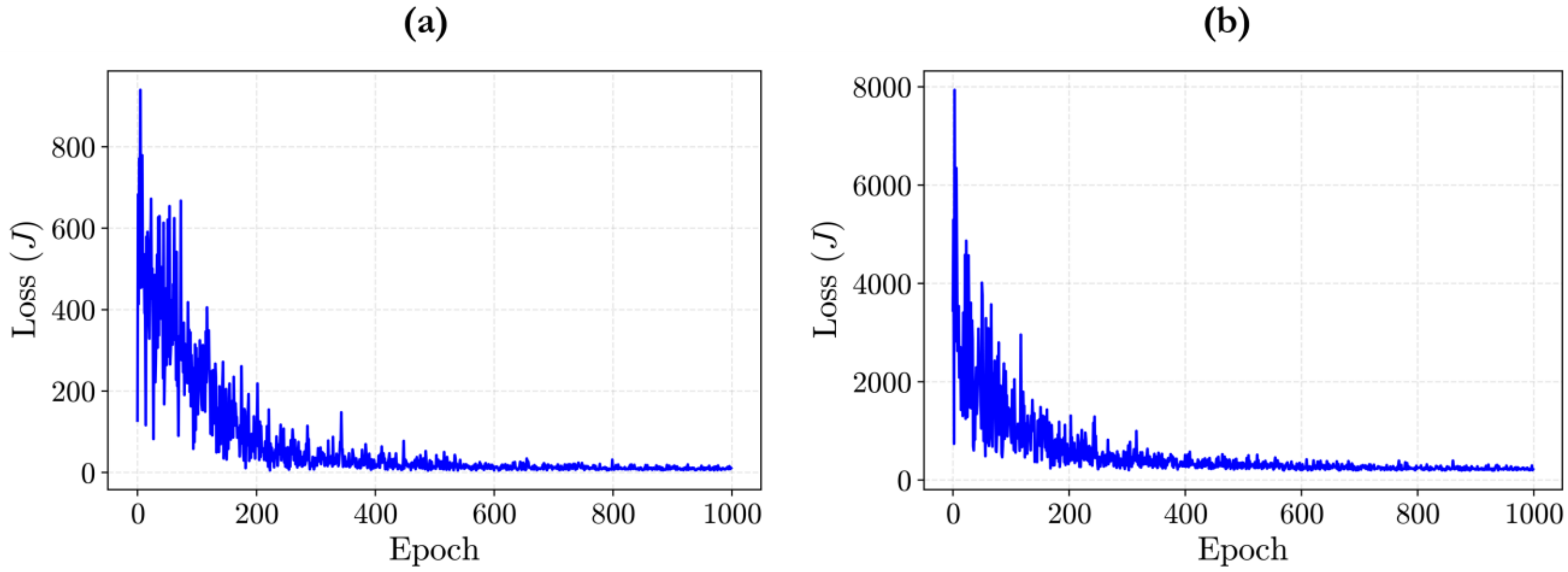

**Fig. A.1.** VI loss trajectories corresponding to $N_{top} = 5$ for **(a)** Approach 1 and **(b)** Approach 2.

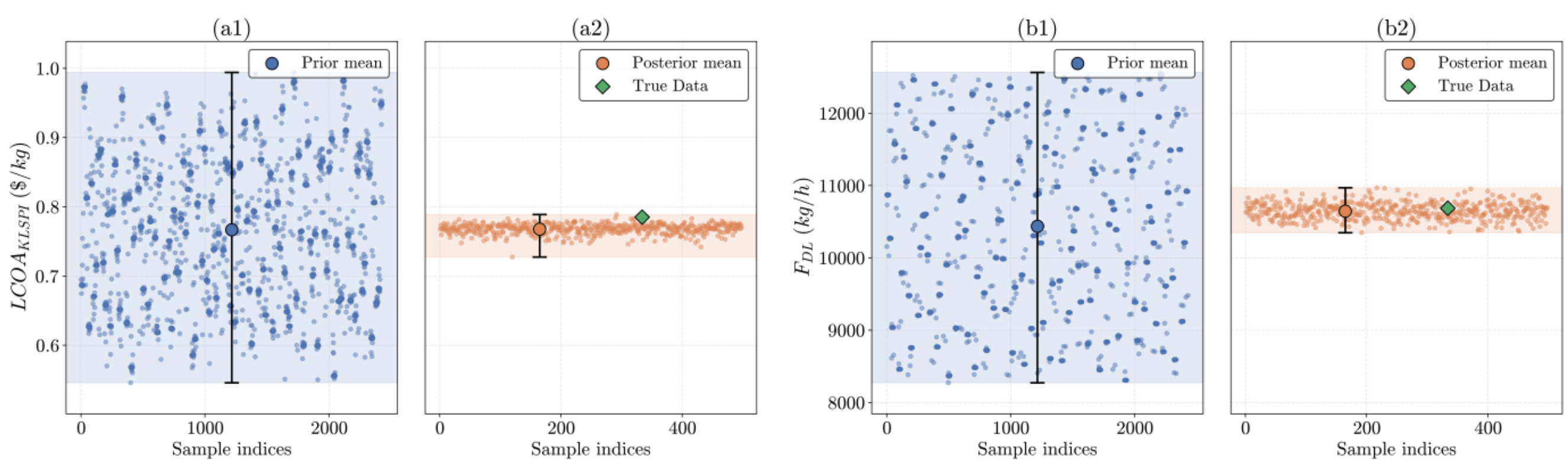

**Fig. A.2**: Prior and posterior predictive results of **(a1)-(a2)** economic (i.e., levelized cost of KL-SPI bioadhesive) and **(b1)-(b2)** process (i.e., D-lignin mass flow rate) outputs for Approach 1 with $N_{top} = 5$.

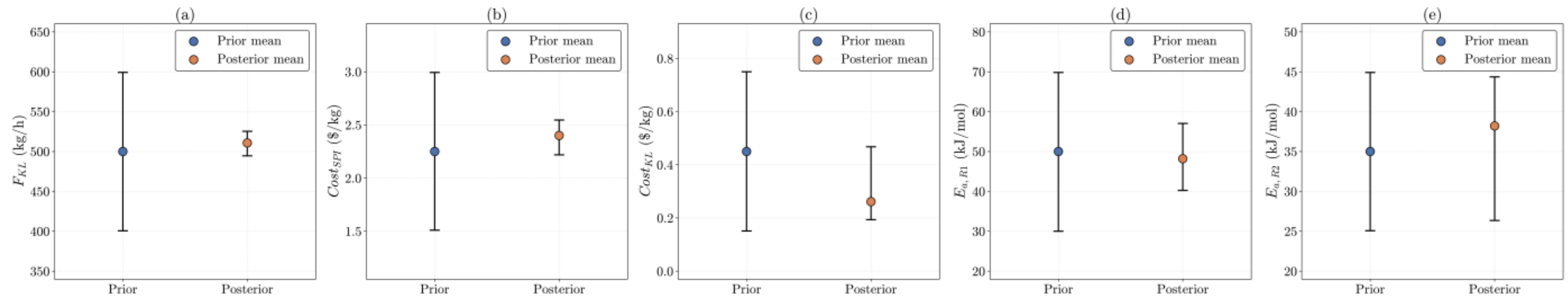

**Fig. A.3**: Prior and posterior means and variances of (a) KL mass flow rate, (b) SPI cost, (c) KL cost, (d) BCD reaction 1 activation energy, and (e) BCD reaction 2 activation energy for Approach 1.

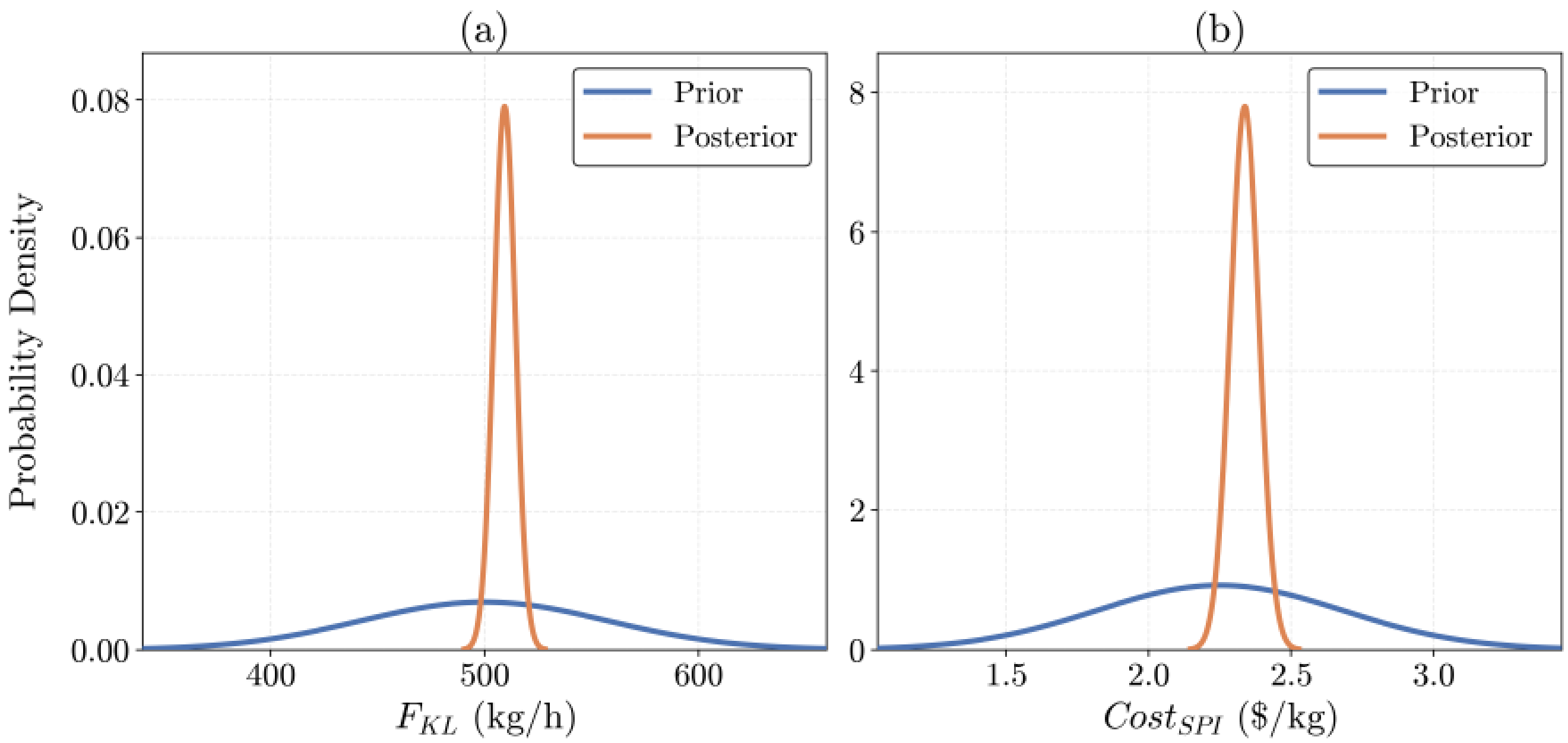

**Fig. A.4.** Prior and posterior distributions of **(a)** KL mass flow rate and **(b)** SPI cost for Approach 1 with $N_{top}$ =2.

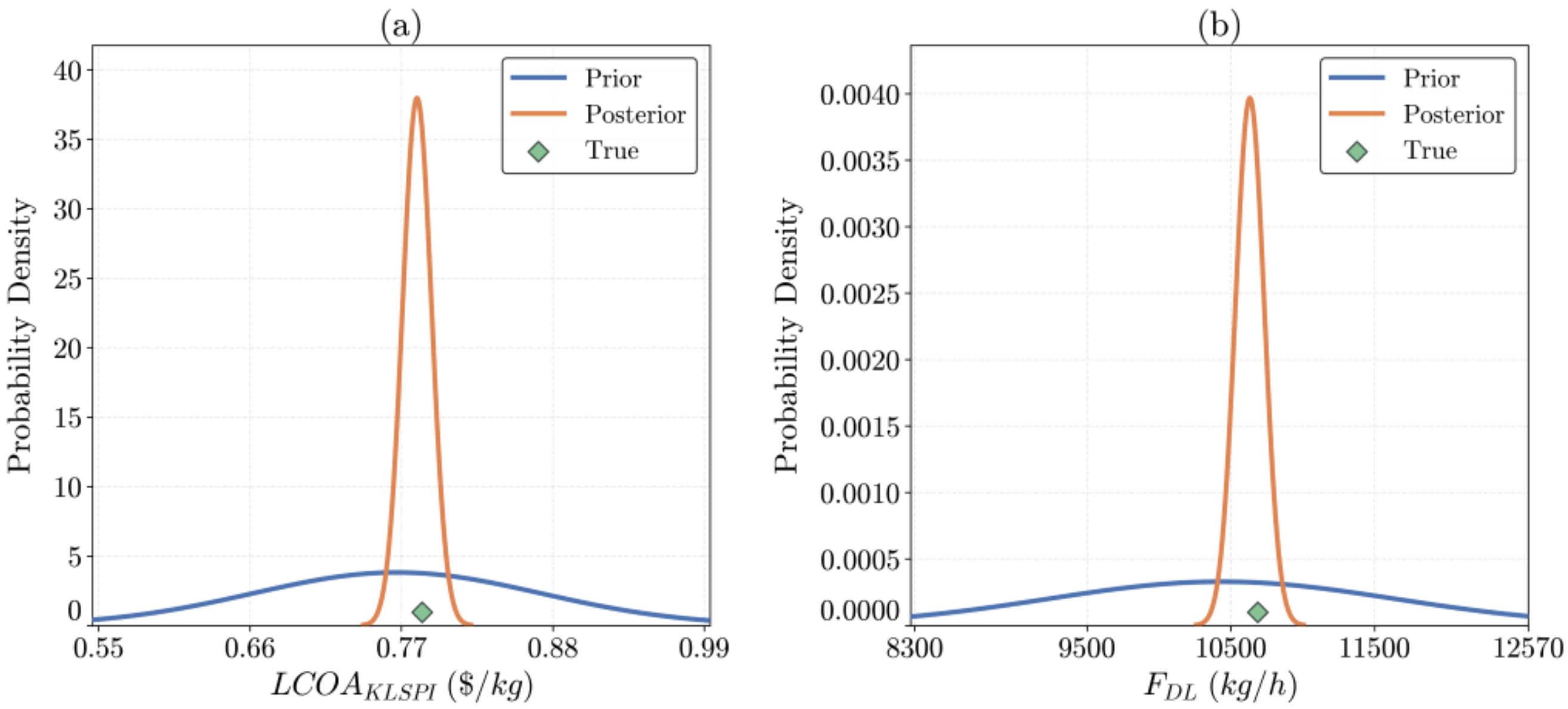

**Fig. A.5.** Prior and posterior predictive distributions of **(a)** economic (i.e., levelized cost of KL-SPI bioadhesive) and **(b)** process (i.e., D-lignin mass flow rate) outputs for Approach 1 with $N_{top}$ = 2. The horizontal limits indicate the range of prior predictive output results obtained from Sobol GSA.

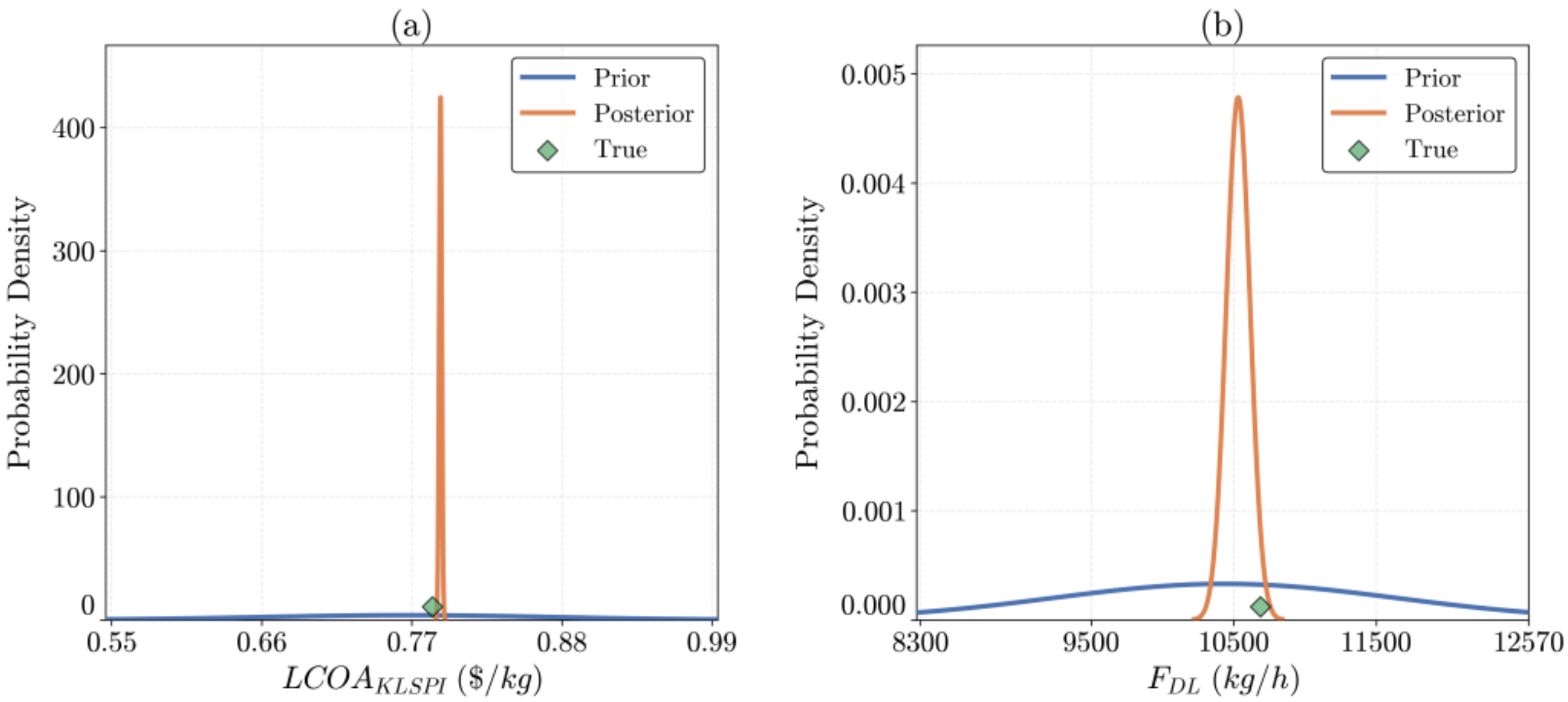

**Fig. A.6.** Prior and posterior predictive distributions of (a) economic (i.e., levelized cost of KL-SPI bioadhesive) and (b) process (i.e., D-lignin mass flow rate) outputs for Approach 1 with $N_{top} = 1$. The horizontal limits indicate the range of prior predictive output results obtained from Sobol GSA.

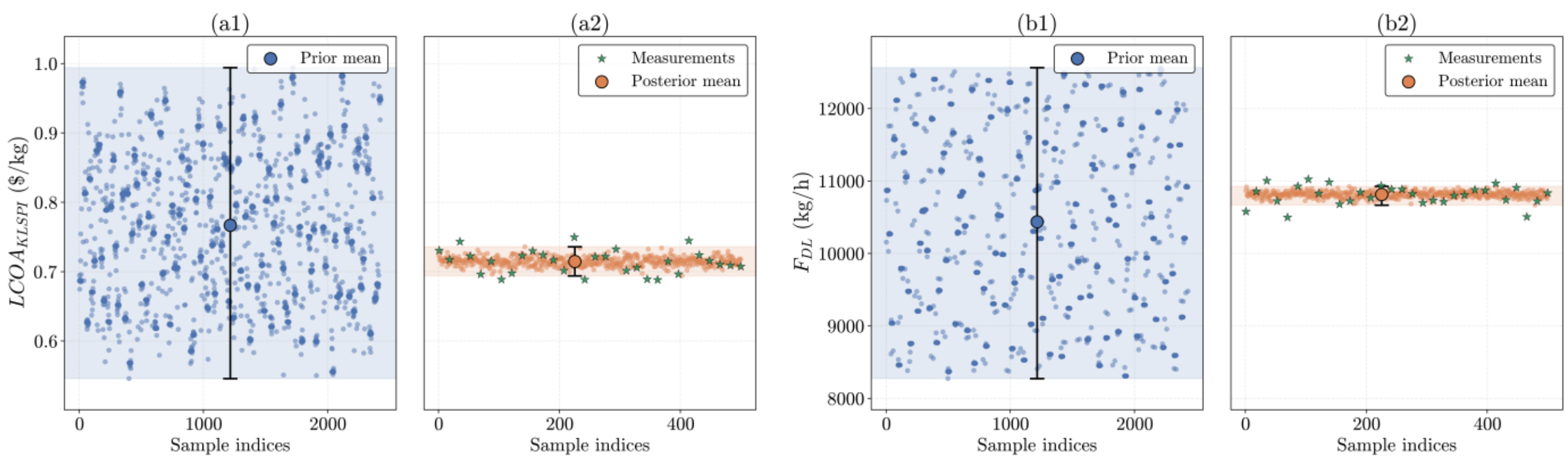

**Fig. A.7.** Prior and posterior predictive results of **(a1)-(a2)** economic (i.e., levelized cost of KL-SPI bioadhesive) and **(b1)-(b2)** process (i.e., D-lignin mass flow rate) outputs for Approach 2 with $N_{top} = 5$.

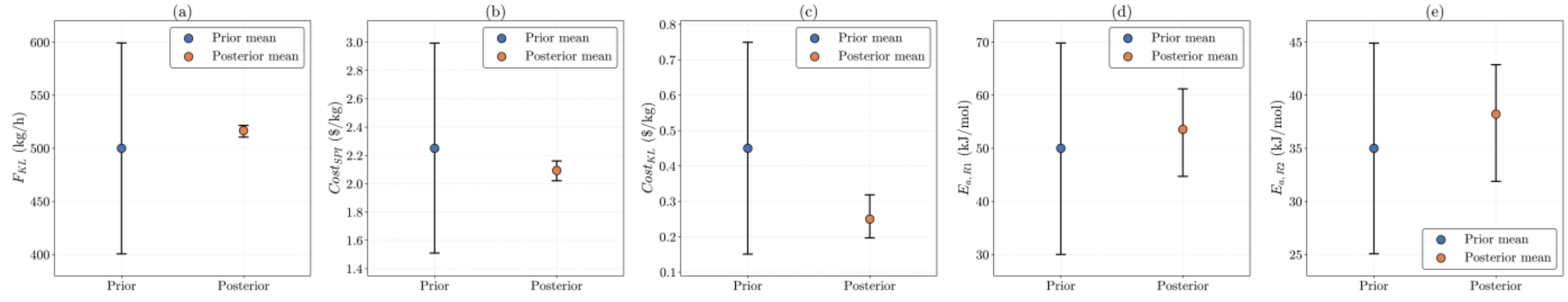

**Fig. A.8.** Prior and posterior means and variances of (a) KL mass flow rate, (b) SPI cost, (c) KL cost, (d) BCD reaction 1 activation energy, and (e) BCD reaction 2 activation energy for Approach 2.

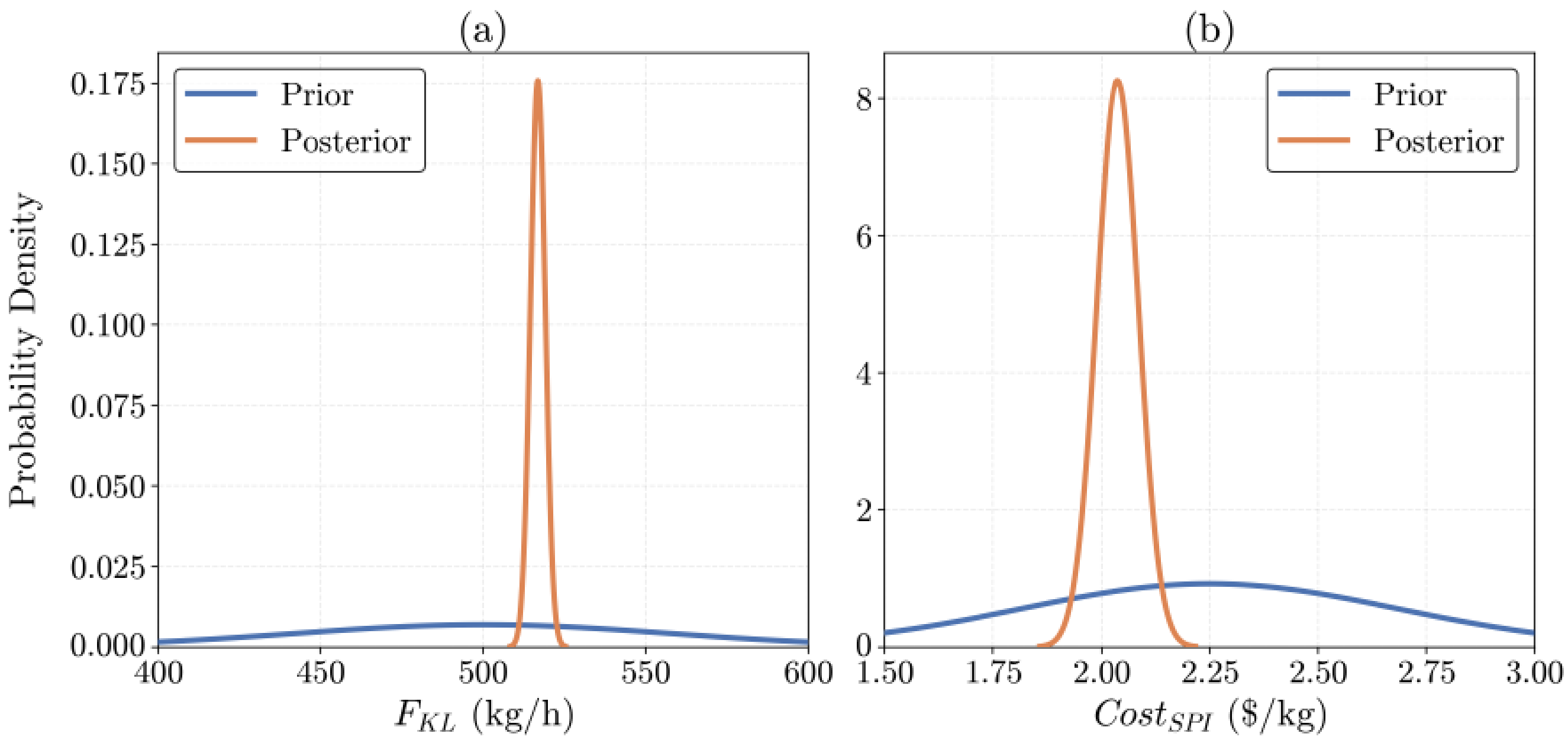

**Fig. A.9.** Prior and posterior distributions of (a) KL mass flow rate and (b) SPI cost for Approach 2 with $N_{top} = 2$.

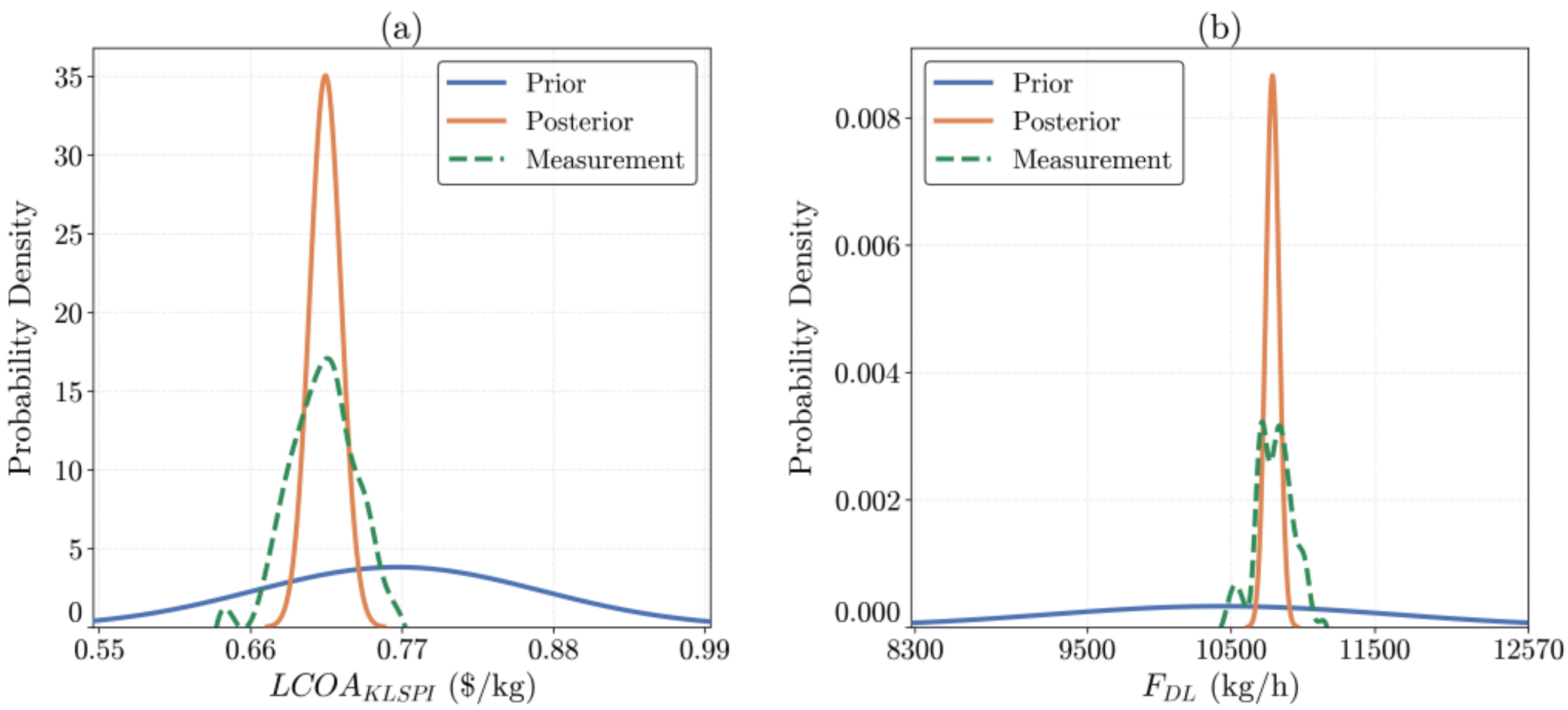

**Fig. A.10.** Prior and posterior predictive distributions of (a) economic (i.e., levelized cost of KL-SPI bioadhesive) and (b) process (i.e., D-lignin mass flow rate) outputs, along with the distribution of synthetic measured data used for VI in Approach 2 with $N_{top} = 2$. The horizontal limits indicate the range of prior predictive output results obtained from Sobol GSA.

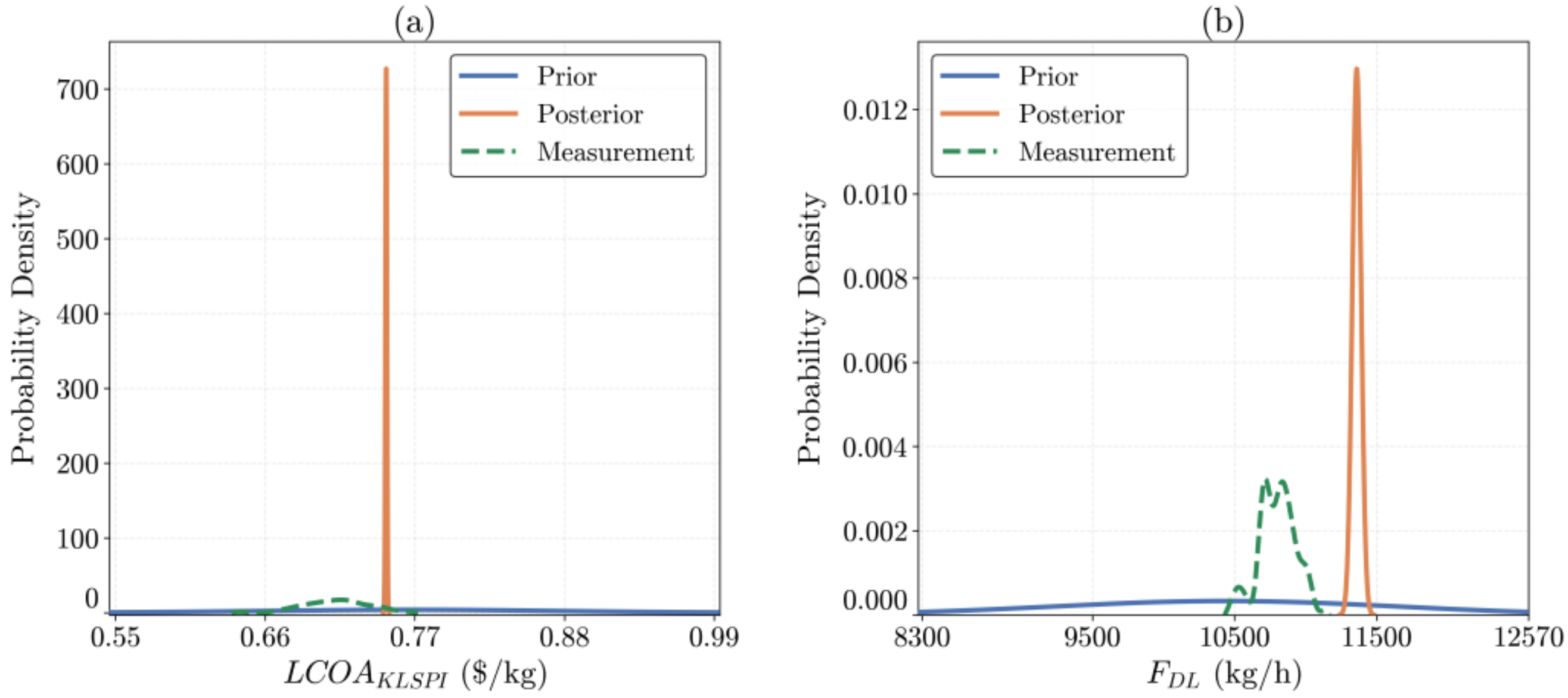

**Fig. A.11.** Prior and posterior predictive distributions of (a) economic (i.e., levelized cost of KL-SPI bioadhesive) and (b) process (i.e., D-lignin mass flow rate) outputs, along with the distribution of synthetic measured data used for VI in Approach 2 with $N_{top} = 1$. The horizontal limits indicate the range of prior predictive output results obtained from Sobol GSA.